\documentclass{ieeetmlcn}
\usepackage{cite}
\usepackage{amsmath,amssymb,amsfonts,amsthm,mathrsfs,bm}
\usepackage{graphicx,color}
\usepackage{subfigure}
\usepackage{textcomp}
\usepackage{xcolor}
\usepackage{hyperref}
\hypersetup{hidelinks=true}
\usepackage{epstopdf}
\usepackage{multirow}
\def\BibTeX{{\rm B\kern-.05em{\sc i\kern-.025em b}\kern-.08em
    T\kern-.1667em\lower.7ex\hbox{E}\kern-.125emX}}
\AtBeginDocument{\definecolor{tmlcncolor}{cmyk}{0.93,0.59,0.15,0.02}\definecolor{NavyBlue}{RGB}{0,86,125}}

\def\OJlogo{\vspace{-4pt}\includegraphics[height=18pt]{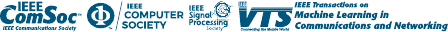}}
\def\seclogo{\vspace{10pt}\includegraphics[height=18pt]{new_logo-eps-converted-to.pdf}}

\def\authorrefmark#1{\ensuremath{^{\textbf{#1}}}}

\usepackage[linesnumbered,ruled,vlined]{algorithm2e}
\usepackage{algpseudocode}

\newtheorem{theorem}{Theorem}

\newtheorem{assumption}{Assumption}
\newtheorem{definition}{Definition}

\begin{document}
\receiveddate{13, May, 2025}
\reviseddate{04, Jul, 2025}
\accepteddate{08 Sep, 2025}

\markboth{Analysis and Optimization of Wireless Multimodal Federated Learning on Modal Heterogeneity}{Han {et al.}}

\title{Analysis and Optimization of Wireless Multimodal Federated Learning on Modal Heterogeneity}

\author{Xuefeng Han\authorrefmark{1}, Wen Chen\authorrefmark{1}, Jun Li\authorrefmark{2}, Ming Ding\authorrefmark{3}, Qingqing Wu\authorrefmark{1}, \par Kang Wei\authorrefmark{4}, Xiumei Deng\authorrefmark{5}, Yumeng Shao\authorrefmark{6}, Qiong Wu\authorrefmark{7}}
\affil{Department of Electronic Engineering, Shanghai Jiao Tong University, Minhang 200240, China}
\affil{School of Information Science and Engineering, Southeast University, Nanjing, 210096, China}
\affil{Data61, CSIRO, Sydney, NSW 2015, Australia}
\affil{School of Cyber Science and Engineering, Southeast University, Nanjing, 211189, China}
\affil{Pillar of Information Systems Technology and Design
Singapore University of Technology and Design, Singapore}
\affil{School of Electrical and Optical Engineering, Nanjing University of Science and Technology, Nanjing 210094, China}
\affil{School of Internet of Things Engineering, Jiangnan University, Wuxi 214122, China}
\corresp{Corresponding author: Wen Chen (email: wenchen@sjtu.edu.cn).}
\authornote{This work is supported by NSFC Key Project 62531015 and Shanghai Kewei 24DP1500500. This work is supported in part by the Key Technologies R\&D Program of Jiangsu (Prospective and Key Technologies for Industry) under Grants BE2023022 and BE2023022-2, in part by National Natural Science Foundation of China (NSFC) under Grant 62471204, and in part by Major Natural Science Foundation of the Higher Education Institutions of Jiangsu Province under Grant 24KJA510003. This work is supported by NSFC 62371289 and NSFC 62331022. This work is supported by the Fundamental Research Funds for the Central Universities (Grant No. 2242025K30025). }

\begin{abstract}
Multimodal federated learning (MFL) is a distributed framework for training multimodal models without uploading local multimodal data of clients, thereby effectively protecting client privacy. However, multimodal data is commonly heterogeneous across diverse clients, where each client possesses only a subset of all modalities, renders conventional analysis results and optimization methods in unimodal federated learning inapplicable. In addition, fixed latency demand and limited communication bandwidth pose significant challenges for deploying MFL in wireless scenarios. To optimize the wireless MFL performance on modal heterogeneity, this paper proposes a joint client scheduling and bandwidth allocation (JCSBA) algorithm based on a  decision-level fusion architecture with adding a unimodal loss function. Specifically, with the decision results, the unimodal loss functions are added to both the training objective and local update loss functions to accelerate multimodal convergence and improve unimodal performance. To characterize MFL performance, we derive a closed-form upper bound related to client and modality scheduling and minimize the derived bound under the latency, energy, and bandwidth constraints through JCSBA. Experimental results on multimodal datasets demonstrate that the JCSBA algorithm improves the multimodal accuracy and the unimodal accuracy by 4.06\% and 2.73\%, respectively, compared to conventional algorithms.
\end{abstract}

\begin{IEEEkeywords}
Multimodal federated learning, bandwidth allocation, client scheduling, modality heterogeneity
\end{IEEEkeywords}


\maketitle

\section{INTRODUCTION}

\IEEEPARstart{I}{ntegrated} artificial intelligence (AI) and communication emerge as a usage scenario of the sixth generation mobile network (6G). On the one hand, AI techniques can be leveraged to promote existing communication services. On the other hand, advanced communication systems can help to provide better training and inference of AI models. Nevertheless, the wide application of AI sparks growing public concerns about potential data privacy leakage. As a distributed machine learning paradigm, federated learning (FL) was first proposed by Google in \cite{Google}. Over time, FL has further applications in blockchain \cite{block_chain}, differential privacy \cite{differential_privacy}, large language models \cite{FL+LLM}, \textcolor{black}{and can integrate with federated fine-tuning \cite{add1} and over-the-air computation \cite{add2}.} Unlike the conventional machine learning centrally trains models, FL enables clients to update local models on local datasets while only uploading updated local models to a server for global aggregation. Hence, this distributed paradigm eliminates the need to upload raw data and preserves client privacy. The process of FL is divided into a series of communication rounds, each consisting of model broadcasting, local updates, model uploading, and global aggregation. Due to the extensive coverage and massive access ability, wireless networks serve as the backbone for model broadcasting and model uploading. However, most wireless clients are placed in complex and diverse physical environments, so that data collected by clients is commonly multimodal, such as camera equipment and recording equipment in meetings or various sensors in a factory. To effectively process such multimodal inputs, existing wireless FL on unimodal data has to integrate with multimodal learning, which can effectively improve the model performance \cite{MFL_recommend}.
\par
A typical multimodal learning architecture consists of unimodal submodels and a fusion process. The key distinction between multimodal learning and unimodal learning lies in the fusion process, which can be categorized into feature-level fusion \cite{feature1}, decision-level fusion \cite{decision1}, and hybrid fusion \cite{hybrid1}. In the feature-level fusion, all unimodal representation vectors are flattened into an input vector for the fusion model. Since the fusion model requires a fixed input dimension, the feature-level fusion cannot adapt to datasets with missing modalities. However, the decision-level fusion only sums or averages unimodal decision results to obtain the final decision result, and such a simple parameter-free operator supports all possible multimodal combinations.
\par
In multimodal learning, different convergence speeds among modalities lead to modality imbalance \cite{modal_rebalance}. Consequently, inadequate training on unconverged modalities degrades multimodal performance, while excessive training on converged modalities results in overfitting. When multimodal learning is integrated with FL, multimodal federated learning (MFL) further suffers from modality heterogeneity since clients collect only a part of all modalities constrained by their sensors \cite{survey2}. Such modal heterogeneity challenges the MFL architecture and exacerbates the modality imbalance. Once MFL performs in the wireless networks, computation and communication are constrained by fixed latency in each communication round, where a strictly fair bandwidth allocation and client scheduling strategy may lead to transmission failures\cite{limited_resource}. Furthermore, the modality heterogeneity introduces significant variations in latency, posing challenges to the existing algorithms about bandwidth allocation and client scheduling for wireless unimodal FL. After all, limited bandwidth cannot support all clients participating in MFL, and part of clients participating may cause no update for a missing modality. As thus, a comprehensive wireless MFL architecture considering modality heterogeneity and a corresponding algorithm about bandwidth allocation and client scheduling are essential for enhancing MFL performance.

\subsection{Related Works}
Most optimization works of FL in wireless networks focused on unimodal learning. \cite{maximize_data} scheduled clients and allocated bandwidth to enhance FL performance. Furthermore, \cite{epoch} adjusted the local epoch number according to the latency constraint, while \cite{reinforcement_learning} adopted reinforcement learning to schedule clients with rewards about the latency of FL. Based on the blockchain assisted FL, \cite{FL_blockchain} also scheduled clients to maximize the long-term time average training data. And \cite{FL_energy} further considered the stochastic energy arrivals of clients and the corresponding optimization of wireless FL performance. Moreover, \cite{quantization} utilized the model quantization to reduce the energy consumption. Despite ignoring multimodal data, \cite{maximize_data, importance_aware, epoch, reinforcement_learning, FL_blockchain, FL_energy, quantization} provided a thorough thought for wireless FL optimization.
\par
Now we turn our eyes to MFL. To address the modality heterogeneity among clients, some works reconstructed the representations of the missing modalities. Based on the similarity between available modalities and missing modalities, \cite{distribution} obtained the representation distribution of the missing modalities. Adding the reconstruction loss on available modalities of clients, \cite{contrastive_loss} trained autoencoders, and reconstructed the representations of the missing modalities. As for a more complicated situation, where a client has partial unimodal data and partial multimodal data, a transformer was utilized in \cite{modal_transformer} for reconstruction. Despite providing complete methods for modality reconstruction in \cite{distribution, contrastive_loss, modal_transformer}, it is undesirable for wireless clients constrained by latency and energy to support additional tons of training.
\par
Keeping local multimodal models with different architectures for model heterogeneity, knowledge distillation (KD) helps to aggregate knowledge from clients. In the global aggregation, \cite{distillation} aggregated knowledge about outputs of local multimodal models on the proxy dataset to train the global multimodal model. When KD also performs in local updates, local multimodal models can utilize the corresponding modalities of the proxy dataset to learn the output of the global multimodal model. As thus, \cite{contrast_representation} exchanged global knowledge and local knowledge between the server and clients. And \cite{select_client} further selected clients with different modality combinations at fixed ratios to maintain balanced modality participation. Nevertheless, collecting the proxy dataset reflecting the global distribution is significantly challenging. Moreover, sharing the proxy dataset among clients introduces additional communication overhead and privacy leakage risks.
\par
Apart from additional training and datasets, some works modified the multimodal fusion process. Assuming representations of unimodal submodels with the same dimensions, \cite{regular_term} summed up the representations as the input of the fusion model. And \cite{dropout} further adopted modality dropout with a probability for multimodal clients. Considering the restrictive assumption of the same dimensions for representation, \cite{select} employed the decision-level fusion, and selected both modalities and clients based on their training performance, communication overhead, and recency. \cite{wise_and_fusion} divided the training process of MFL into federated unimodal learning and federated fusion learning to train unimodal submodels and the fusion model, respectively. Admittedly, \cite{regular_term, dropout, select, wise_and_fusion} provided good ideas to tackle the modality heterogeneity, avoiding extra computation and communication overhead. Unfortunately, none of these prior works take the wireless scenario of MFL into consideration, nor concrete limitations of bandwidth, latency, or energy.
\par
\begin{figure*}[t]
  \centering
  \includegraphics[width=.75\textwidth]{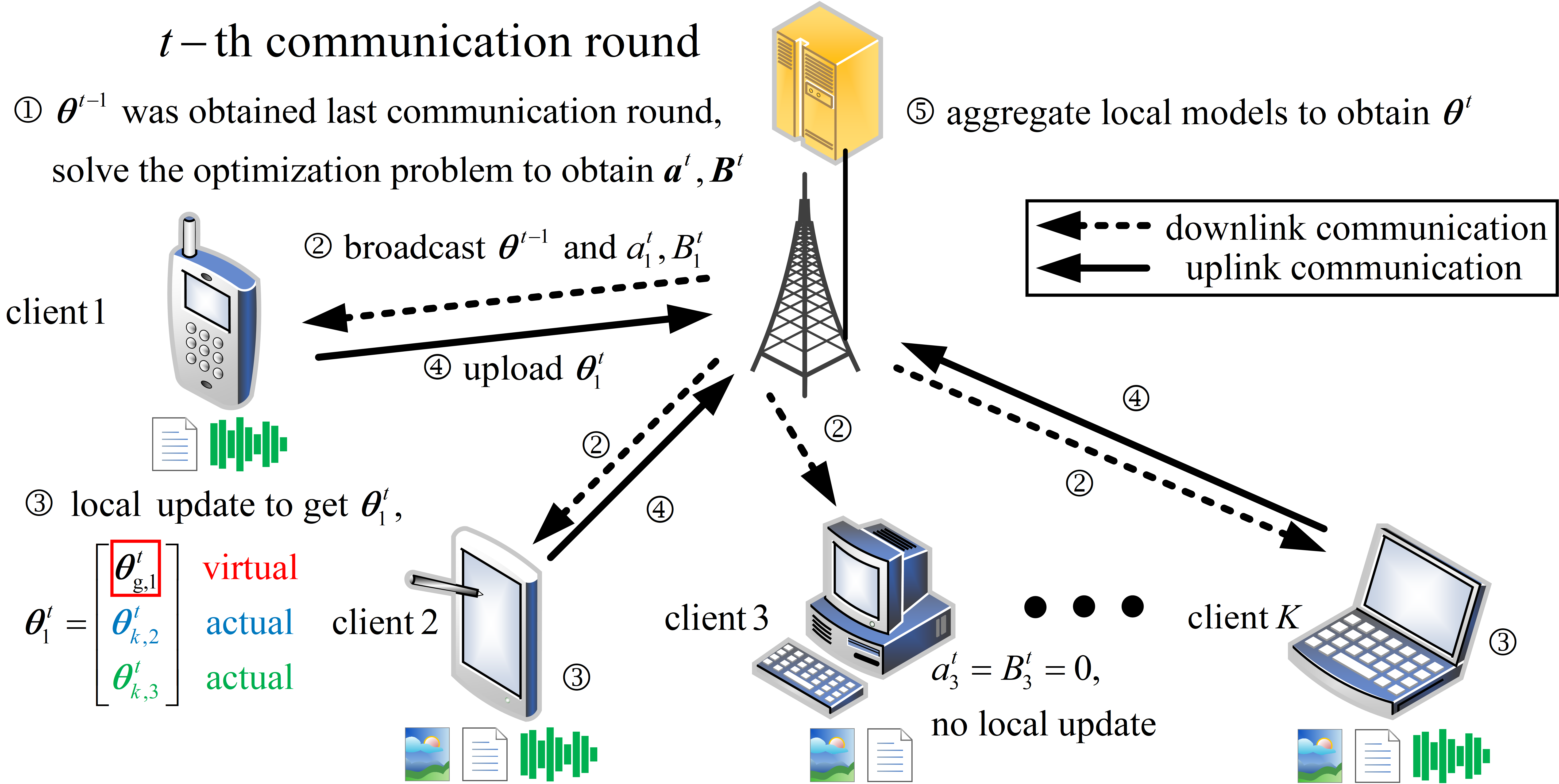}
  \caption{Each communication round in wireless MFL consists of 5 steps. For client 1 without the image modality, $\bm \theta_{{\rm g},1}^t$ is a virtual global unimodal submodel which is actually not uploaded. Client 3 is not scheduled in the $t$-th communication round, so that there is no local update nor uploading on client 3.}
  \label{figure: architecture}
\end{figure*}
Some recent works extended to wireless MFL. While \cite{iot} proposed a hybrid multimodal federated learning in the industrial Internet of things, \cite{incentive} tackled a resource management problem with integrated sensing, computing, and communication to enhance MFL performance. However, \cite{iot, incentive} both ignored the modality heterogeneity in MFL. In a nutshell, most existing works have doubtful performance in the wireless MFL on modality heterogeneity. To the best of our knowledge, the algorithm for scheduling clients with modality heterogeneity and allocating bandwidth in the wireless MFL has not been studied in the literature.

\subsection{Contributions}
This paper proposes a wireless MFL framework addressing the modality heterogeneity among clients. \textcolor{black}{The key focus of the proposed framework lies in adding unimodal loss functions based on the decision-level fusion, along with the joint client scheduling and bandwidth allocation (JCSBA) algorithm to optimize MFL performance. This integration introduces no additional computational overhead for enhancing unimodal performance, which is crucial for clients with limited computational and wireless resources. To the best of our knowledge, this is the first work to optimize unimodal-assisted, decision-level MFL on modal heterogeneity in wireless networks.} The main contributions are summarized as follows.
\par
\begin{itemize}
\item For clients with modality heterogeneity, we propose a wireless MFL framework based on decision-level fusion. With the output result of each unimodal submodel, an unimodal loss term is added to the objective function of wireless MFL, and a corresponding gradient term of the unimodal loss is applied in local updates to improve unimodal and multimodal performance.  The framework supports the local updates for any multimodal combination of unimodal submodels and aggregates local unimodal submodels to obtain the global multimodal model comprising all global unimodal submodels.
\item Based on the proposed wireless MFL framework, the latency and energy consumption of clients with modality heterogeneity during the communication and computation are modeled. An MFL performance optimization is then formulated under bandwidth, latency, and energy constraints to schedule clients and allocate bandwidth.
\item To characterize the performance of MFL, a closed-form upper bound on the decrease of the loss function is derived based on the scheduling results of modalities and clients. Its form suggests that scheduling should prioritize clients with unconverged modalities. Using this upper bound as the transformed objective function, we rely on Lyapunov optimization techniques, the Tammer decomposition method, Karush-Kuhn-Tucker (KKT) conditions, and the immune algorithm to solve the optimization problem about JCSBA.
\item Simulation results on multimodal datasets validate that our JCSBA algorithm achieves the fastest convergence, the best performance, and the least energy consumption. Specifically, compared to the Selection algorithm \cite{select_client}, the Dropout algorithm \cite{dropout}, and the other conventional algorithms, the proposed JCSBA algorithm improves the multimodal accuracy by 1.67\%, 3.35\%, and 4.06\%, respectively, while also enhancing the unimodal accuracy by 2.65\%, 1.35\%, and 2.73\%.
\end{itemize}
\par
The rest of this article is organized as follows. The wireless MFL framework is introduced in Section II. The corresponding physical models are described in Section III. In Section IV, a closed-form upper bound is derived to characterize MFL performance. Section V then solves the optimization problem of wireless MFL. Section VI shows the experimental results, and Section VII draws a conclusion.

\section{Multimodal Federated Learning Framework over Wireless Networks}
\subsection{Multimodal Federated Learning Process}
Consider a cellular network including $K$ clients and one server connected to a base station (BS) to accomplish an MFL task as Fig. \ref{figure: architecture}. $M$ modalities among all clients constitute $\mathcal M = \{1, 2, \cdots, M\}$. Modalities of client $k$ constitute $\mathcal M_k \subseteq \mathcal M$ and $| \mathcal M_k|$ denotes the modal number of client $k$. It is assumed that the data of available modalities is aligned in each client. Hence, the dataset of client $k$ is expressed by $\mathcal D_k = \{ (\{ \bm x_{k,m,j} | m \in \mathcal M_k \}, y_{k,j}) | j = 1, 2, \cdots, D_k \}$, where $D_k$ is the size of the dataset, $\bm x_{k,m,j}$ is the feature vector of modality $m$ in sample $j$, and $y_{k,j}$ is the label of sample $j$.
\par
\begin{figure}[t]
  \centering
  \includegraphics[width=.48\textwidth]{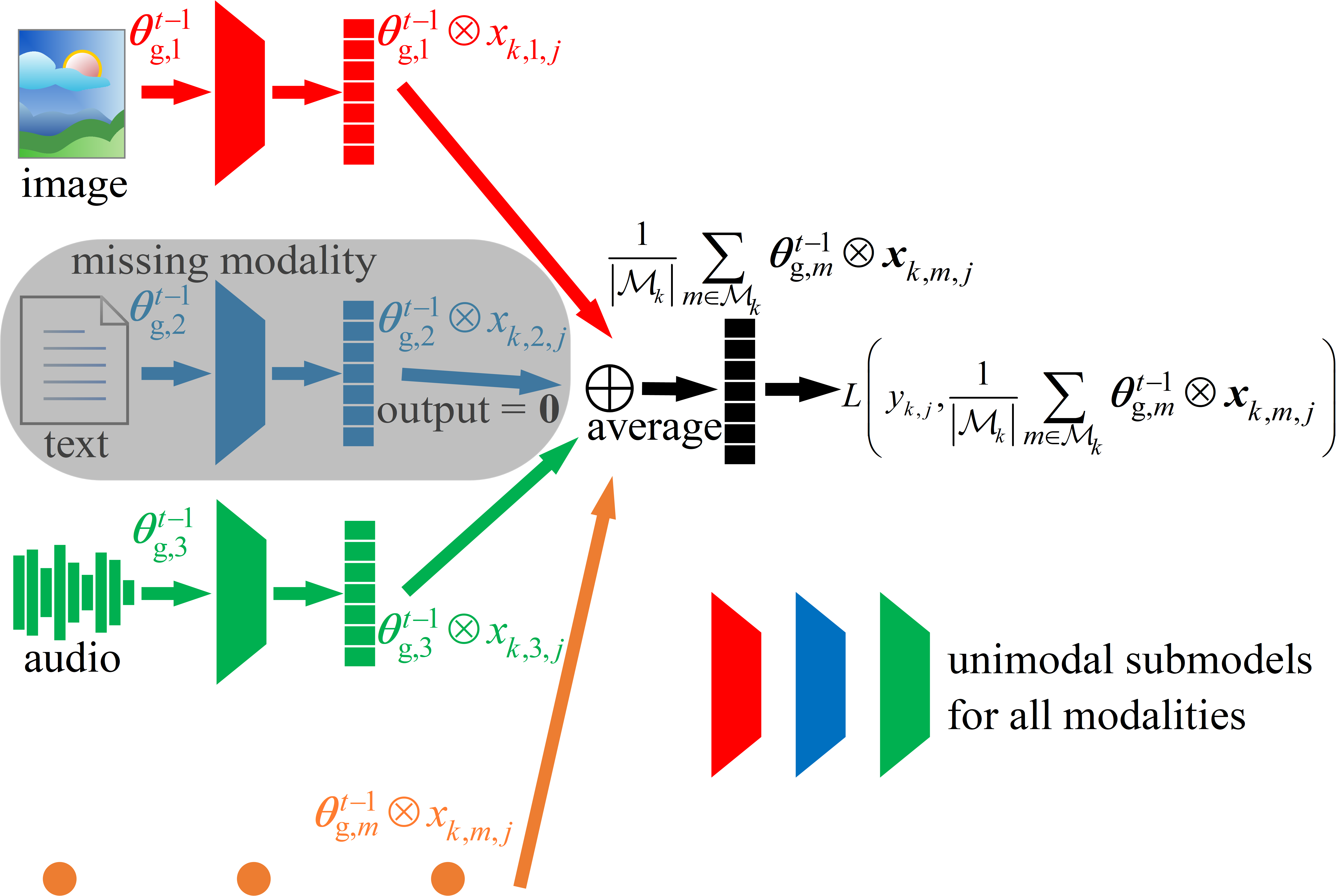}
  \caption{Decision-level fusion is employed in wireless MFL. The multimodal model computes the loss function on sample $j$ of client $k$. Since the text modality is missing, its output is set to zero.}
  \label{figure: model}
\end{figure}

The whole progress of MFL is divided into $T$ communication rounds. In the $t$-th communication round, the global multimodal model $\bm \theta^{t-1}$ is broadcast to all clients. After receiving $\bm \theta^{t-1}$, each participating client needs to execute the local update on its dataset. Remarkably, there are three different modal fusion methods for multimodal learning. Nevertheless, to adapt to the modal heterogeneity among clients and avoid redundant burden in computation and communication, a multimodal model with the decision-level fusion in Fig. \ref{figure: model} is utilized in our MFL. $\bm \theta^{t-1} = [(\bm \theta^{t-1}_{{\rm g}, 1})^\top, (\bm \theta^{t-1}_{{\rm g}, 2})^\top, \cdots, (\bm \theta^{t-1}_{{\rm g}, M})^\top]^\top$ is divided into $M$ global unimodal submodels, and each one is corresponding to a modality. Specifically, the output of the $\bm \theta_{{\rm g}, m}^{t-1}$ correspond to modality $m$ for input $\bm x_{k,m,j}$ is denoted by $\bm \theta_{{\rm g}, m}^{t-1} \otimes \bm x_{k,m,j}$. If some modalities are missing, the corresponding outputs will be set to $\bm 0$. And the fusion method is the average of outputs with equal weights of modalities \cite{modal_blend}. As such, the multimodal loss of client $k$ is
\begin{equation}
F_k(\bm \theta^{t-1}) = \frac{1}{D_k} \sum_{j \in \mathcal D_k} L \Big( y_{k,j} , \frac{1}{|\mathcal M_k|} \sum_{m \in \mathcal M_k} \bm \theta^{t-1}_{{\rm g}, m} \otimes \bm x_{k,m,j} \Big),
\label{equation: multimodal loss}
\end{equation}
where $L(\cdot)$ determines the formula of the loss function. In the classification task, $L(\cdot)$ is normally the cross entropy error between the output after ${\rm SoftMax}(\cdot)$ and the one-hot label. And in the regression task, $L(\cdot)$ is the mean square error between the output and the label.
\par
Apart from the multimodal local loss function in (\ref{equation: multimodal loss}), MFL should consider unimodal performance, which adapts to other unimodal clients employing unimodal submodels for inference. Moreover, \cite{dropout, select, wise_and_fusion} validate that multimodal performance can benefit from the training of unimodal submodels. The unimodal loss function helps to balance different convergence speeds among modalities, which also accelerates the convergence of the multimodal model \cite{crossmodal_infiltration}. As thus, we obtain the sum of all unimodal loss functions in client $k$ is
\begin{equation}
G_k(\bm \theta^{t-1}) = \sum_{j\in \mathcal D_k} \sum_{m=1}^M G_k(\bm \theta_{{\rm g}, m}^{t-1}).
\label{equation: unimodal loss}
\end{equation}
From (\ref{equation: unimodal loss}), we notice that client $k$ only has a part of the modalities. For available modality $m \in \mathcal M_k$, the unimodal loss function is computed by
\begin{equation}
G_k(\bm \theta_{{\rm g}, m}^{t-1}) = \frac{v_m}{D_k} \sum_{j \in \mathcal D_k} L(y_{k,j}, \bm \theta_{{\rm g}, m}^{t-1} \otimes \bm x_{k,m,j}),
\label{equation: existing modal loss}
\end{equation}
where $v_m$ is a pre-set modal weight. As for missing modality $m \notin \mathcal M_k$, the unimodal loss function cannot be simply set to zero since a zero loss leads to a zero gradient, leading to biased global aggregation. Thus, the unimodal loss function is defined by the global unimodal loss, whose gradient is the global unimodal gradient, i.e.,
$
G_k(\bm \theta_{{\rm g}, m}^{t-1}) \triangleq G(\bm \theta_{{\rm g}, m}^{t-1}),
$ for $m \notin \mathcal M_k$.
Summing up the multimodal loss and the unimodal loss, the local loss is
\begin{equation}
H_k(\bm \theta^{t-1}) = F_k(\bm \theta^{t-1}) + G_k(\bm \theta^{t-1}). \\
\label{equation: local loss}
\end{equation}
\textcolor{black}{In fact, adding the unimodal loss in (\ref{equation: local loss}) introduces no additional computational overhead. This is because the outputs of unimodal submodels, i.e., $\bm \theta_{{\rm g}, m}^{t-1} \otimes \bm x_{k,m,j}$, are obtained during the computation of the averaged multimodal output. Only the cross entropy between the unimodal output and the one-hot label is computed in (\ref{equation: existing modal loss}), and its computational cost is negligible compared to the complexity of forward and backward propagation in neural networks.}
\par
In the local update, we assume that there is only one epoch in each communication round, and the batch gradient descent (BGD) is adopted. Thus the local gradient is
\begin{equation}
\nabla H_k(\bm \theta^{t-1}) = \nabla F_k(\bm \theta^{t-1}) + \nabla G_k(\bm \theta^{t-1}).
\label{equation: local gradient}
\end{equation}
Just like the local multimodal model, the local multimodal gradient $\nabla H_k(\bm\theta^{t-1}) = [\nabla H_k^\top(\bm\theta^{t-1}_{{\rm g}, 1}),$ $ \nabla H_k^\top(\bm\theta^{t-1}_{{\rm g}, 2}), $ $ \cdots,  \nabla H_k^\top(\bm\theta^{t-1}_{{\rm g}, M})]^\top$ is also divided into $M$ unimodal subgradients. For $m \in \mathcal M_k$, the local unimodal subgradient with respect to $\bm \theta_m^{t-1}$ is computed by
\begin{align}
& \nabla H_k(\bm \theta_m^{t-1}) = \frac{v_m}{D_k} \sum_{j \in \mathcal D_k} \nabla L(y_{k,j}, \bm \theta_{{\rm g}, m}^t \otimes \bm x_{k,m,j}) \label{equation: existing local gradient} \\
& \qquad\quad + \frac{1}{D_k}\sum_{j\in \mathcal D_k} \frac{\partial  L \Big( y_{k,j} , \frac{1}{|\mathcal M_k|} \sum_{i \in \mathcal M_k} \bm \theta^{t-1}_{{\rm g}, i} \otimes \bm x_{k,i,j} \Big)}{\partial \bm \theta_{{\rm g}, m}^{t-1}}. \notag
\end{align}
For $m \notin \mathcal M_k$, similarly, the local unimodal subgradient is defined by
$
\nabla H_k(\bm\theta_m^{t-1}) \triangleq \nabla H(\bm\theta_m^{t-1}).
$
Finally, the local update is expressed by
\begin{equation}
\bm \theta_k^t =
\begin{bmatrix} \bm \theta_{k,1}^t \\ \bm \theta_{k,2}^t \\ \vdots \\ \bm \theta_{k,M}^t \end{bmatrix}
=
\begin{bmatrix} \bm \theta_{{\rm g}, 1}^{t-1} \\ \bm \theta_{{\rm g}, 2}^{t-1} \\ \vdots \\ \bm \theta_{{\rm g}, M}^{t-1} \end{bmatrix}
- \eta
\begin{bmatrix} \nabla H_k(\bm \theta_{{\rm g}, 1}^{t-1}) \\ \nabla H_k(\bm \theta_{{\rm g}, 2}^{t-1}) \\ \vdots \\ \nabla H_k(\bm \theta_{{\rm g}, M}^{t-1}) \end{bmatrix}.
\label{equation: local update matrix}
\end{equation}
Despite the global loss and the global gradient for missing modalities, client $k$ neither needs nor updates them in the practical training since we have the identical equation $\bm \theta_{k,m}^t = \bm\theta_{{\rm g}, m}^{t-1} - \eta \nabla H(\bm\theta_{{\rm g}, m}^{t-1}) = \bm\theta_{{\rm g}, m}^t$ for $m \notin \mathcal M_k$. The above definitions for missing modalities are aimed at the same shape of local multimodal models and a concise aggregation formula.
\par
Now we turn to global aggregation. The global loss function, a weighted sum of all local loss functions, is first written by
\vspace{-3mm}
\begin{equation}
\begin{aligned}
H(\bm \theta^{t-1}) & = \sum_{k=1}^K w_k H_k(\bm \theta^{t-1}) \\
& = \sum_{k=1}^K w_k F_k(\bm \theta^{t-1}) + \sum_{k=1}^K w_k G_k(\bm \theta_{{\rm g}, m}^{t-1}) \\
& = F(\bm \theta^{t-1}) + G(\bm \theta^{t-1}),
\end{aligned}
\label{equation: global loss}
\end{equation}
where $w_k \triangleq \frac{D_k}{ \sum_i^K D_i}$. The global unimodal subgradient for modality $m$ is
\begin{align}
\nabla H(\bm \theta_{{\rm g}, m}^{t-1}) & = \sum_{k \in \mathcal K_m} w_k \nabla H_k(\bm \theta_{{\rm g}, m}^{t-1}) + \sum_{k \notin \mathcal K_m} w_k \nabla H(\bm \theta_{{\rm g}, m}^{t-1}) \notag \\
& = \sum_{k \in \mathcal K_m} \bar w_{k,m} \nabla H_k(\bm \theta_{{\rm g}, m}^{t-1}) , \label{equation: global gradient}
\end{align}
where \textcolor{black}{$\bar w_{k,m} \triangleq \frac{w_k}{\sum_{i \in \mathcal K_m} w_i}$ denotes the unified aggregation weight.} In (\ref{equation: global gradient}), we can observe that $\sum_{k \notin \mathcal K_m} w_k \nabla H(\bm\theta^{t-1}_{{\rm g}, m})$  can be transposed to the left side, and $\nabla H(\bm\theta^{t-1}_{{\rm g}, m})$ is computed by the sum of $\nabla H_k(\bm\theta^{t-1}_{{\rm g}, m})$ for $k\in \mathcal K_m$.
\par
As for the global aggregation in the server, the global unimodal submodel of modality $m$ is
\begin{align}
\bm \theta^t_{{\rm g}, m} & = \sum_{k=1}^K w_k \bm \theta_{k,m}^t = \sum_{k \in \mathcal K_m} w_k \bm \theta_{k,m}^t + \sum_{k\notin\mathcal K_m} w_k \bm \theta_{{\rm g}, m}^t \notag \\
& = \sum_{k \in \mathcal K_m} \bar w_{k,m} \bm \theta^t_{k,m} = \bm \theta_{{\rm g}, m}^{t-1} - \eta \nabla H(\bm\theta_{{\rm g}, m}^{t-1}). \label{equation: global aggregation}
\end{align}
It is observed that the update direction of the global unimodal submodel in (\ref{equation: global aggregation}) is the direction of the negative global unimodal subgradient in (\ref{equation: global gradient}), demonstrating that our MFL framework for clients with missing modalities is unbiased.
\par
After $T$ communication rounds, the loss function of the global multimodal model $\bm\theta^T$ is expected to get minimum. Namely, the aim of MFL is
\begin{equation}
\min_{\bm \theta^T} \Big( H(\bm \theta^T) = F(\bm \theta^T) + G(\bm \theta^T) \Big).
\label{equation: initial aim}
\end{equation}
\vspace{-5mm}
\subsection{Partial Participation of Wireless Clients}
Due to the limited communication bandwidth, only a part of the clients are scheduled. A participation vector is $\bm a^t = [a_1^t, a_2^t, \cdots, a_K^t] \in \{ 0, 1 \}^K$, where $a_k^t = 1$ indicates that client $k$ participates in the $t$-th communication round, and vice versa. $\bm B^t = [B_1^t, B_2^t, \cdots, B_K^t]$ denotes the corresponding allocation bandwidth vector. All participating clients constitute $\mathcal K^t = \{ k\in \mathcal K | a_k^t = 1 \}$, and participating clients with $m$ modal data constitute $\mathcal K^t_m = \{k \in \mathcal K^t | m \in \mathcal M_k \}$.
\par
Different from (\ref{equation: global aggregation}), the global aggregation of $\bm\theta^t_m$ is among $\mathcal K^t$, and the server may encounter a situation that none of the participating clients have modality $m$. Hence $\bm\theta_{{\rm g}, m}^t$ remains $\bm\theta_{{\rm g}, m}^{t-1}$ for $m \notin \mathcal M^t= \{i \in \mathcal M | \mathcal K^t_i \neq \varnothing \}$. As for $m \in \mathcal M^t $, (\ref{equation: global aggregation}) is modified into
\begin{equation}
\bm \theta_{{\rm g}, m}^t = \bm \theta^{t-1}_{{\rm g}, m} - \eta \sum_{k \in \mathcal K^t_m} w^t_{k,m} \nabla H_k(\bm\theta_{{\rm g}, m}^{t-1}).
\label{equation: participation local update}
\end{equation}
\textcolor{black}{where the participated aggregation weight $w_{k,m}^t$ is defined by $w_{k,m}^t \triangleq \frac{D_k}{\sum_{i \in \mathcal K^t_m} D_i}$.}
\par
Herein, the framework of MFL for heterogeneous modalities and partial client participation is developed. The JCSBA algorithm is presented in \textbf{Algorithm \ref{algorithm: JCSBA}}.
\begin{algorithm}
  \label{algorithm: JCSBA}
  \caption{JCSBA algorithm for wireless MFL}
  \KwOut{multimodal global model\ $\bm \theta^T$}
  Initialize $\bm \theta^0$ in the server\;
  \For{$t = 1, 2, \cdots, T$}
  {
    Server solves the optimization problem to obtain $\bm a^t, \bm B^t$ and broadcasts $\bm \theta^{t-1}$ to $\mathcal K$\;
    \For{$k \in \mathcal K^t$ in parallel}
    {
        Update local model $\bm \theta_k^t$ according to (\ref{equation: local update matrix})\;
        Upload updated the local multimodal model $\bm \theta_k^t$ with bandwidth $B_k^t$ to the server\;
    }
    Update the global multimodal model according to (\ref{equation: global aggregation})\;
  }
  \Return $\bm \theta^T$\;
\end{algorithm}

\section{System Model and Problem Formulation}
\textcolor{black}{Based on the wireless MFL framework, both the communication overhead and the computation overhead are modeled, and corresponding constraints are considered to formulate an optimization problem.}
\subsection{Communication Model}
Communication progress in MFL can be divided into two parts: downlink communication and uplink communication. The downlink communication relies on the server broadcasting. Benefiting from the large downlink transmitting power on the server and sufficient downlink bandwidth, the latency of the downlink communication can be ignored compared to the latency of the uplink communication \cite{downlink}. Similarly, there is no concern about energy consumption on the server, due to its abundant energy supply.
\par
For the uplink communication, clients upload their models by means of frequency division multiple access (FDMA). According to Shannon's equation, the uplink rate of client $k \in \mathcal K^t$ in the $t$-th communication round is
\begin{equation}
r_k^t = B_k^t \log_2(1 + \frac{p h _k^t}{B_k^t N_0}),
\label{equation: uplink rate}
\end{equation}
where $B_k^t$ is the bandwidth allocated to client $k$, $p$ is the uplink transmitting power, and $N_0$ is noise power spectral density. $h_k^t$ is the channel gain influenced by large scale fading, small scale fading, antenna gain, and other gains.
Furthermore, there is a total bandwidth allocated to participating clients in FDMA, which can be expressed by
\begin{equation}
\sum_{k = 1}^{K} a_i^t B_k^t \leq B^{\max}.
\label{equation: uplink bandwidth}
\end{equation}
Due to modal heterogeneity in MFL, different clients actually upload different local unimodal submodels. Since local unimodal submodels among clients for a modality have the same structure, $\ell_m$ can denote the bit length of the local unimodal submodel for modality $m$. And the uplink latency of client $k$ in the $t$-th communication round is expressed by
\begin{equation}
\tau_k^{t, {\rm com}} = \frac{\sum_{m \in \mathcal M} \ell_m}{r_k^t} = \frac{\Gamma_m}{r_k^t}.
\label{equation: uplink latency}
\end{equation}
Consequently, the uplink energy consumption is
\begin{equation}
e_k^{t, {\rm com}} = p \tau_k^{t, {\rm com}}.
\label{equation: uplink energy}
\end{equation}
\subsection{Computation Model}
Due to enough computational resources in the server, the latency of global aggregation can be ignored compared to the latency of local updates. It is assumed that all clients have the same computational abilities, i.e., frequency $\nu$ of central processing units (CPUs) and energy consumption coefficient $\alpha$. Similar to $\ell_m$ in the uplink communication, $\beta_m$ denotes the number of CPU cycles computing a sample on modality $m$. Based on the structure of the model in Fig. 2, numbers of CPU cycles on different modalities are additive \cite{d2d}. And the fusion of modal outputs needs $(|\mathcal M_k| - 1) \beta_0$ CPU cycles since the outputs from different modalities have the same dimensions. Hence, the computation latency of client $k$ is
\begin{equation}
\tau_k^{{\rm cmp}} = \frac{D_k \left[\sum_{m \in \mathcal M_k} (\beta_m + \beta_0) -  \beta_0 \right] }{f} = \frac{D_k \Phi_k}{f}.
\label{equation: computation latency}
\end{equation}
The computation energy consumption is then written by
\begin{equation}
e_k^{{\rm cmp}} = \alpha D_k f^2 \Big[\sum_{m \in \mathcal M_k} \!\! (\beta_m + \beta_0) -  \beta_0 \Big] = \alpha D_k f^2 \Phi_k.
\label{equation: compucation energy}
\end{equation}
\subsection{Optimal Problem}
For each participating client in the $t$-th communication round, the server can not wait without the limitation. Hence, client $k \in \mathcal K^t$ should satisfy the latency constraint as
\begin{equation}
\tau_k^{t, {\rm com}} + \tau_k^{{\rm cmp}} \leq \tau^{\max},
\label{equation: latency constraint}
\end{equation}
where $\tau^{\max}$ is the maximal latency in each communication round.
\par
Apart from the latency, the energy consumption of clients is also limited. Different from the latency, residual energy in the current communication round can be stored for later computation and communication. Hence, the residual energy queue in the $t$-th communication round is
$
q^t_k = E^{\rm add} - a_k^t (e_k^{t, {\rm com}} + e_k^{{\rm cmp}}),
$
where $E^{\rm add}$ is allocated energy for MFL in each communication round. After $T$ communication rounds, the energy consumption cannot exceed the storage, that is
\begin{equation}
\sum_{t=1}^T q_k^t \geq 0.
\label{equation: energy constraint}
\end{equation}
\par
Under the constraints in (\ref{equation: uplink bandwidth}), (\ref{equation: latency constraint}) and (\ref{equation: energy constraint}), (\ref{equation: initial aim}) serves as the objective function, thus the optimization problem is
\par
\begin{equation}
\begin{aligned}
\textbf{ P1: } &  \min_{\{ (\bm a^t, \bm B^t) |t \} } \lim_{T \rightarrow + \infty} H(\bm\theta^T), \\
{\rm s.t.} \enspace & \textbf{C1: } a_k^t \in \{0, 1\}, \quad \forall k \in \mathcal{K}, \\
& \textbf{C2: } B_k^t \geq 0, \quad \forall k \in \mathcal K, \\
& \textbf{C3: } \sum_{k = 1}^{K} a_i^t B_k^t \leq B^{\max}, \\
& \textbf{C4: } a_k^t(\tau_k^{t, {\rm com}} + \tau_k^{{\rm cmp}}) \leq \tau^{\max}, \quad \forall k \in \mathcal K, \\
& \textbf{C5: } \lim_{T \rightarrow + \infty} \frac{1}{T}\sum_{t=1}^T q_k^t \geq 0, \enspace \forall k \in \mathcal{K}.
\end{aligned}
\label{equation: optimal problem}
\end{equation}
where $T \!\! \rightarrow \!\!+ \infty$ means that the communication round number ensures the global multimodal model converges. As $H(\bm \theta^T)$ cannot be obtained before training, and \textbf{C5} is a long-term constraint , we then analyse $H(\bm \theta^T)$ to characterize MFL performance in Section IV, and transform \textbf{P1} into an instantaneous problem to solve in Section V.

\section{MFL Performance Analysis}
Motivated by \cite{bound}, we can derive an upper bound of $H(\bm \theta^T)$ and guide our subsequent optimization. \textcolor{black}{For ease of reference, we include the variables during the MFL performance analysis of this chapter in Table \ref{table: variable}.}

\begin{table}[ht]\color{black}
  \centering
  \caption{\textcolor{black}{List of Main Notations in MFL Performance Analysis}}
\begin{tabular}{c|c}
\hline \hline
\textbf{Notation} & \textbf{Description} \\
\hline
$\bm \theta^t$ & Global multimodal model \\
\hline
$\bm \theta^t_{{\rm g}, m}$ & Global unimodal model of modality $m$ \\
\hline
$\bm \psi^t$ & Virtual auxiliary parameter \\
\hline
$\bar w_{k,m}$ & Unified aggregation weight \\
\hline
$w^t_{k,m}$ & Participated aggregation weight \\
\hline
$H(\bm\theta^t)$ & Global multimodal loss function \\
\hline
$\nabla H(\bm\theta^t)$ & Global multimodal gradient \\
\hline
$\nabla H(\bm \theta^t_{{\rm g}, m})$ & Global unimodal subgradient of modality $m$ \\
\hline
$\nabla H_k(\bm \theta^t_{{\rm g}, m})$ & Local\! unimodal\! subgradient\! of\! modality\! $m$\! and\! client\! $k$ \\
\hline
$\gamma$ & Smooth parameter of the global loss function \\
\hline
$\rho$ & Lipschitz parameter of the global loss function \\
\hline
\multirow{2}{*}{$\zeta_m^t$} & Local unimodal subgradient upper bound of \\
&  modality $m$ \\
\hline
\multirow{2}{*}{$\delta^t_{k,m}$} & Subgradient divergence upper bound of \\
& modality $m$ and client $k$ \\
\hline \hline
\end{tabular}
\label{table: variable}
\end{table}

\subsection{Definition and Assumption}

As the premise, we give \textbf{Assumption \ref{assumption: smooth}}-\textbf{\ref{assumption: divergence}} corresponding to $H(\bm \theta)$ as follows.
\begin{assumption}
The global loss function is $\gamma$-smooth, i.e., $\| \nabla H(\bm\theta) -\! \nabla H(\bm\varphi) \| \! \leq \gamma \| \bm\theta - \bm\varphi \|, \forall \bm \theta, \bm\phi$.
\label{assumption: smooth}
\end{assumption}
\begin{assumption}
The global loss function is $\rho$-Lipschitz continuous, i.e., $| H(\bm\theta) - H(\bm\varphi) | \leq \rho \| \bm\theta - \bm\varphi \|, \forall \bm \theta, \bm\phi$.
\label{assumption: Lipschitz}
\end{assumption}
\begin{assumption}
The norm of the local unimodal subgradient is bounded by $\| \nabla H(\bm\theta_{{\rm g}, m}^{t-1}) \| \leq \zeta_m^{t-1}$.
\label{assumption: bounded gradient}
\end{assumption}
\begin{assumption}
The norm of the difference between the global unimodal subgradient and the local unimodal subgradient is bounded by $\| \nabla H_k(\bm\theta_{{\rm g}, m}^{t-1}) - \nabla H(\bm\theta_{{\rm g}, m}^{t-1}) \| \leq \delta_{k,m}^{t-1}$.
\label{assumption: divergence}
\end{assumption}
\textcolor{black}{Inspired by \cite{iot, modal_blend}, the characteristics of the multimodal model are described in \textbf{Assumption 1} and \textbf{Assumption 2}, whereas \textbf{Assumption 3} and \textbf{Assumption 4} outline the training processes of different unimodal models.} In addition, we define a virtual auxiliary parameter to server a bridge between the global multimodal models in the adjacent communication rounds.
\begin{definition}
The virtual auxiliary parameter $\bm \psi_m^t$ is updated by local multimodal gradients of all clients, i.e., $ \bm\psi_m^t \triangleq \bm\theta_{{\rm g}, m}^{t-1} - \eta \sum_{k=1}^K \bar w_{k,m} \nabla H_k(\bm\theta_{{\rm g}, m}^{t-1}) = \bm\theta_{{\rm g}, m}^{t-1} - \eta \nabla H(\bm\theta_{{\rm g}, m}^{t-1}).$
\label{definition: auxiliary parameter}
\end{definition}
\subsection{Main Results}

\begin{theorem}
The difference between $H(\bm\theta^t)$ and $H(\bm\psi^t)$ is bounded by
\begin{equation}
H(\bm\theta^t) - H(\bm\psi^t) \leq \textcolor{black}{\eta \rho \sqrt{A_1^t + A_2^t}} , \label{equation: theorem1}
\end{equation}
\textcolor{black}{where $A_1^t \triangleq \sum_{m \notin \mathcal M^t} (\zeta_m^{t-1})^2$, $A_2^t \triangleq \sum_{ m \in \mathcal M^t } [2 (1 - \sum_{k \in \mathcal K_m} a_k^t \bar w_{k,m}) \sum_{k \in \mathcal K_m}  (w^t_{k,m} \!\! + \bar w_{k,m} \!\! - 2a_k^t \bar w_{k,m}) (\delta_{k,m}^{t-1})^2 ] $ for concise expression.}
\label{theorem1}
\end{theorem}
\vspace{-3mm}
\begin{IEEEproof}
Please refer to the detailed proof in Appendix \ref{proof}.
\end{IEEEproof}

In \textbf{Theorem 1}, all clients participation makes the whole term equal 0, which is consistent with reality. We can also see that $M$ modalities act together in the upper bound. \textcolor{black}{If $\zeta_m^{t-1} \gg \delta_{k,m}^{t-1}$, modality $m$ will contribute more significantly to the bound value in $A_1^t$ than $A_2^t$. Hence, the scheduling decision will emphasize modality $m$ to avoid large bound values. In contrast, if $\zeta_m^{t-1} \ll \delta_{k,m}^{t-1}$, modality $m$ will have a greater impact on the bound value in $A_2^t$ than in $A_1^t$. And the scheduling decision will consider suspending training on modality $m$ in favor of clients with other modalities. A large $\delta_{k,m}^{t-1}$ indicates that $\bm\theta^{t-1}_{{\rm g}, m}$ is necessary and preferred during client scheduling.}

\begin{theorem}
The descend of the loss function in a communication round is bounded by
\begin{align}
& H(\bm\theta^t) - H(\bm\theta^{t-1}) \\
& \leq  - \frac{2\eta - \gamma \eta^2}{2} \sum_{m=1}^M \| \nabla H(\bm \theta^{t-1}_{{\rm g}, m}) \|^2 + \eta \rho \sqrt{ A_1^t + A_2^t } . \notag
\end{align}
\label{theorem2}
\end{theorem}
\vspace{-7mm}
\begin{IEEEproof}
Firstly, the difference between a communication round is taken into consideration as
\begin{equation}
H(\bm\theta^t) - H(\bm\theta^{t-1}) = (H(\bm\theta^t) - H(\bm\psi^t)) + (H(\bm\psi^t) - H(\bm\theta^{t-1})).
\label{proof: theorem2_1}
\end{equation}
Now that $H(\bm\theta^t) - H(\bm\psi^t)$ is bounded in \textbf{Theorem \ref{theorem1}}, we can concentrate on $H(\bm\psi^t) - H(\bm\theta^{t-1})$. According to the property of $\gamma$-smooth in \textbf{Assumption \ref{assumption: smooth}}, we have
\begin{equation}
H(\bm\psi^t) - H(\bm\theta^{t-1}) \leq \langle H(\bm\theta^{t-1}), \bm\psi^t - \bm\theta^{t-1} \rangle + \frac{\gamma}{2} \| \bm\psi^t - \bm\theta^{t-1} \|^2.
\label{proof: theorem2_2}
\end{equation}
The square norm term can be expanded into a sum over $M$ modalities as
\begin{align}
\frac{\gamma}{2} \| \bm\psi^t - \bm\theta^{t-1} \|^2 & = \frac{\gamma \eta^2}{2} \| \nabla H(\bm\theta^{t-1}) \|^2 \label{proof: theorem2_3} \\
& = \frac{\gamma \eta^2}{2} \sum_{m=1}^M \| \nabla H(\bm\theta_{{\rm g}, m}^{t-1}) \|^2. \notag
\end{align}
Similarly, the inner product term can be rewritten by
\begin{align}
\langle H(\bm\theta^{t-1}), \bm\psi^t - \bm\theta^{t-1} \rangle & \! = \! \langle \nabla H(\bm \theta^{t-1}), - \eta \nabla H(\bm\theta^{t-1}) \rangle \notag \\
& = -\eta \sum_{m=1}^M \Big\| \nabla H(\bm\theta_{{\rm g}, m}^{t-1}) \Big\|^2. \label{proof: theorem2_4}
\end{align}
Substituting (\ref{proof: theorem2_3}) and (\ref{proof: theorem2_4}) into (\ref{proof: theorem2_2}), we have
\begin{equation}
H(\bm\psi^t) - H(\bm\theta^{t-1})  \leq - \frac{2\eta - \gamma \eta^2}{2} \sum_{m=1}^M \| \nabla H(\bm\theta_{{\rm g}, m}^{t-1}) \|^2. \label{proof: theorem2_5}
\end{equation}
Substituting (\ref{proof: theorem2_5}) into (\ref{proof: theorem2_1}) and applying \textbf{Theorem \ref{theorem1}}, we can prove \textbf{Theorem \ref{theorem2}}.
\end{IEEEproof}
The first term on the right side in \textbf{Theorem \ref{theorem2}} is negative and unrelated to $\bm a^t$. Hence, the descend of the loss function correlated to $\bm a^t$ is only characterized by $\eta\rho\sqrt{A_1^t + A_2^t}$.
\section{Problem Solution}
\begin{figure}[t]
  \centering
  \includegraphics[width=0.49\textwidth]{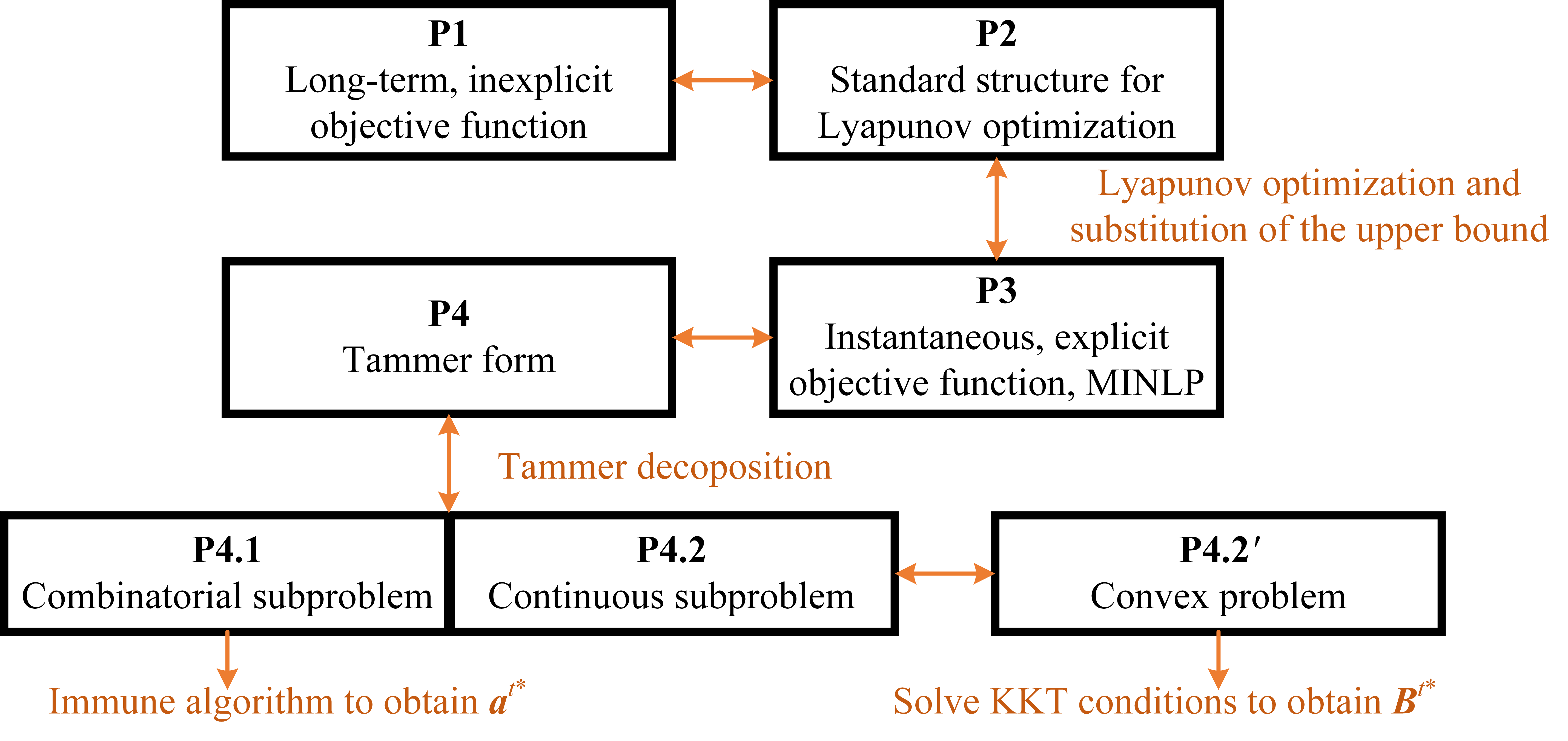}
  \caption{\textcolor{black}{Solving process of P1 to obtain $(\bm a^{t*}, \bm B^{t*})$ within the JCSBA algorithm.}}
  \label{figure: solution}
\end{figure}

With Lyapunov optimization techniques and the derived upper bound, the long-term optimization problem \textbf{P1} is transformed into a series of instantaneous optimization problems across all communication rounds. And each instantaneous optimization problem in a communication round is decomposed into two subproblems which can be individually solved. \textcolor{black}{The solving process of \textbf{P1} is presented in Figure \ref{figure: solution}.}
\subsection{Problem Transformation}
For the long-term constraint \textbf{C5}, we firstly define a virtual queue as
$
Q^{t}_k = \max \{Q^{t-1}_k - q^{t-1}_k , 0\}.
$
Equivalent to \textbf{C5}, the mean stability of the virtual queue is written by
\begin{equation}
\textbf{C5}'\textbf{:} \lim_{T \rightarrow +\infty} \frac{|Q^T_k|}{T} = 0.
\label{equation: stable vritual queue}
\end{equation}
With $H(\bm \theta^T)$ minus a constant term $H(\bm \theta^0)$ acting the new objective function, we obtain a equivalent optimization problem as
\begin{align}
& \textbf{ P2: } \min_{\{ (\bm a^t, \bm B^t) | t \} } \lim_{T \rightarrow +\infty} \sum^T_{t=1} (H(\bm \theta^t) - H(\bm \theta^{t-1})) , \notag \\
& \enspace {\rm s.t.} \quad \textbf{C1} \sim \textbf{C4},\ \textbf{C5}'. \label{equation: bound problem}
\end{align}
\par
\textbf{P2} is a standard structure for Lyapunov optimization. With Lyapunov optimization techniques \cite{Lyapunov_book}, the pertubed Lyapunov function and the Laypunov drift-plus-penalty function are respectively
$
\Delta^t = \frac{1}{2} \sum^K_{k=1} (Q_k^t)^2
$ and
$
\Delta^t_V = \mathbb E \{ (\Delta^{t+1})^2 - (\Delta^{t})^2 + V(H(\bm\theta^t) - H(\bm \theta^{t-1})) \ | \ Q_k^t \}.
$
Substituting the upper bound of \textbf{Theorem \ref{theorem2}}, utilizing $(\max{x, 0})^2 \leq x^2$, and enlarging $(q_k^t)^2$ into a positive constant $C_0$, we can derive an upper bound of $\Delta_V^t$ as
\begin{align}
\Delta_V^t & = \frac{1}{2}\sum^K_{k=1} [(\max \{Q_k^t-q_k^t, 0 \})^2 - (Q_k^t)^2] \label{equation: objective function} \\
& \quad - \frac{2\eta - \gamma \eta^2}{2} \sum_{m=1}^M \| \nabla H(\bm\theta_{{\rm g}, m}^{t-1}) \|^2 + V\eta\rho \sqrt{ A_1^t + A_2^t } \notag \\
& \leq \frac{1}{2}\sum^K_{k=1} [ (q_k^t)^2 - 2Q_k^t q_k^t] - \frac{2\eta - \gamma \eta^2}{2} \sum_{m=1}^M \| \nabla H(\bm\theta_{{\rm g}, m}^{t-1}) \|^2 \notag \\
& \quad + V\eta\rho \sqrt{ A_1^t + A_2^t } \notag \\
& \leq \Big( -\sum^K_{k=1} Q_k^t q_k^t + V\eta\rho \sqrt{A_1^t + A_2^t}  + C_0  \notag \\
& \qquad  - \frac{2\eta - \gamma \eta^2}{2} \sum_{m=1}^M \| \nabla H(\bm\theta_{{\rm g}, m}^{t-1}) \|^2 \Big) \triangleq J_0 (\bm a^t, \bm B^t). \notag
\end{align}
Setting $J_0 (\bm a^t, \bm B^t)$ as the objective function, we can find $C_0 - \frac{2\eta - \gamma \eta^2}{2} \sum_{m=1}^M \| \nabla H(\bm\theta_{{\rm g}, m}^{t-1}) \|^2$ are unrelated to $\bm a^t, \bm B^t$. Omitting the above two terms and $Q_k^tE^{\rm add}$, and substituting $q_k^t = E^{\rm add} - a_k^t (e_k^{t, {\rm com}} + e_k^{{\rm cmp}})$ into $J_0$, we can formulate a transformed optimization problem as
\begin{align}
\textbf{ P3: } & \min_{ \bm a^t, \bm B^t }  \Big( J_1(\bm a^t, \bm B^t) = V\eta\rho \sqrt{A_1^t + A_2^t} - \sum^K_{k=1} Q_k^t q_k^t\Big),  \notag \\
{\rm s.t.} \enspace & \textbf{C1} \sim \textbf{C4} \label{equation: Lyapunov problem}
\end{align}
Despite an instantaneous optimization problem in the $t$-th communication round, \textbf{P3} is still mixed integer nonlinear programming (MINLP), which needs further decomposition to solve.
\par
\subsection{Problem Decomposition}
According to Tammer decomposition \cite{Tammer}, a equivalent optimization problem is
\begin{equation}
\begin{aligned}
\textbf{ P4: } \min_{\bm a^t} \left( \min_{\bm B^t} J_1 (\bm a^t, \bm B^t) \right), \qquad \textbf{ s.t.} \enspace \textbf{C1} \sim \textbf{C4}.
\end{aligned}
\label{equation: Equivalent problem}
\end{equation}
The outer combinatorial subproblem with respect to $\bm a^t$ is
\begin{equation}
\begin{aligned}
\textbf{ P4.1: }  \min_{\bm a^t} J_2(\bm a^t), \qquad
\textbf{ s.t.} \enspace \textbf{C1}, \textbf{C3}, \textbf{C4}.
\end{aligned}
\label{equation: Outer combinatorial problem}
\end{equation}

The objective function of \textbf{P4.1} is defined by $J_2(\bm a^t) \triangleq J_1(\bm a^t, \bm B^{t*})$, where $\bm B^{t*}$ is the optimal point of the inner continuous problem, which can be expressed by
\begin{equation}
\begin{aligned}
\textbf{ P4.2: } \min_{\bm B^t} J_3 (\bm B^t ), \qquad
\textbf{ s.t.} \enspace \textbf{C2} \sim \textbf{C4},
\end{aligned}
\label{equation: Inner continuous problem}
\end{equation}
where $J_3(\bm B^t) = \sum_{k \in \mathcal K^t} \frac{Q_k^t p \Gamma_k} {B_k^t \log_2(1 + \frac{p h_k^t}{B_k^t N_0})}$ comes from $J_1$ omitting constant and terms corresponding to $\bm a^t$ since $\bm a^t$ is fixed in \textbf{P4.2}.
\subsection{Continuous Subproblem}
With the given $\bm a^t$, bandwidth should be allocated to only participating clients. Hence, \textbf{P4.2} is rewritten as
\begin{align}
\textbf{ P4.2}&'\textbf{: } \min_{\bm B^t} \sum_{k \in \mathcal K^t} \frac{Q_k^t p \Gamma_k} {B_k^t \log_2(1 + \frac{p h_k^t}{B_k^t N_0})}, \notag \\
{\rm s.t.} \enspace  & \textbf{C2}'\textbf{:} \ B_k^t \leq 0, \quad \forall k \in \mathcal K^t, \label{equation: Modified continuous problem} \\
& \textbf{C3}'\textbf{:} \ \sum_{k \in \mathcal K^t} B_k^t = B^{\max}, \notag \\
& \textbf{C4}'\textbf{:} \ \frac{\Gamma_k}{B_k^t \log_2(1 + \frac{p h_k^t}{B_k^t N_0})} + \frac{D_k \Phi_k}{f} \leq \tau^{\max},\ \forall k \in \mathcal K^t. \notag
\end{align}
Compared to the bandwidth allocation problem in wirless FL, there are different energy queue lengths and computation latency in $\textbf{P4.2}'$. Fortunately, we can prove that $\textbf{P4.2}'$ is a convex problem and utilize Karush-Kuhn-Tucker (KKT) conditions to obtain the optimal bandwidth vector.
\par
The first-order derivative and the second-order derivative of objective function of $\textbf{P4.2}'$ with respect to $B_k^t$ are expressed by
\begin{equation}
\frac{\partial J_3}{\partial B_k^t} = Q_k^t p \Gamma_k (\ln 2) \frac{ \frac{ph_k^t}{ph_k^t + B_k^tN_0} - \ln(1 + \frac{ph_k^t}{B_k^tN_0})}{(B_k^t)^2 \ln^2 (1 + \frac{p h_k^t}{B_k^t N_0})}
\end{equation}
\begin{align}
\frac{\partial^2 J_3}{(\partial B_k^t)^2} = & \frac{Q_k^t p \Gamma_k (\ln 2)}{(B_k^t)^2 \ln^3(1+\frac{ph_k^t}{B_k^tN_0})} \left[ \frac{p^2 (h_k^t)^2 \ln(1+\frac{ph_k^t}{B_k^tN_0})}{(ph_k^t + B_k^tN_0)^2} \right. \notag \\
& \left. + 2 \left( \ln(1+\frac{ph_k^t}{B_k^tN_0}) - \frac{ph_k^t}{B_k^tN_0 + ph_k^t} \right)^2 \right].
\end{align}
It is obvious that $\frac{\partial^2 J_3}{(\partial B_k^t)^2} > 0$. Similarly, we can prove that the function on the left side of $\textbf{C4}'$ is convex. $\textbf{C2}'$ is a convex set and $\textbf{C3}'$ is a linear constraint. As such, $\textbf{P4.2}'$ is a convex optimization problem. Its Lagrange function is written by
$
{\rm La} = \sum_{k \in \mathcal K^t} \!\! \frac{Q_k^t p \Gamma_k} {B_k^t \log_2(1 + \frac{p h_k^t}{B_k^t N_0})} - \! \sum_{k \in \mathcal K^t} \! \lambda_{2,k} B_k^t + \kappa (B^{\max} \!\! - \!\!\! \sum_{k \in \mathcal K^t} \! B_i^t ) + \sum_{k \in \mathcal K^t}  \lambda_{4,k} ( \frac{\Gamma_k}{B_k^t \log_2(1+\frac{ph_k^t}{B_k^tN_0}) } + \frac{D_k \Phi_k}{f} - \tau^{\max} ).
$
Thus the KKT conditions are given by
\begin{small}
\begin{equation}
\left\{
\begin{aligned}
& \lambda_{2,k}, \lambda_{4,k} \geq 0,\quad \sum_{k \in \mathcal K^t} B_k^t - B^{\max} = 0, \\
& -B_k^t \leq 0, \quad \frac{\Gamma_k}{B_k^t \log_2(1 + \frac{p h_k^t}{B_k^t N_0})} \leq \tau^{\max} - \frac{D_k \Phi_k}{f}, \\
& \lambda_{2,k} B_k^t = 0, \enspace \lambda_{4,k} \big( \frac{\Gamma_k}{B_k^t \log_2(1 + \frac{p h_k^t}{B_k^t N_0})} + \frac{D_k \Phi_k}{f} -\tau^{\max} \big)= 0 , \\
& (Q_k^t p \Gamma_k + \lambda_{4,k} \Gamma_k)\frac{ \frac{ph_k^t}{ph_k^t + B_k^tN_0} - \ln(1 + \frac{ph_k^t}{B_k^tN_0})}{ \frac{1}{\ln2}(B_k^t)^2 \ln^2 (1 + \frac{p h_k^t}{B_k^t N_0})} - \lambda_{2,k} \! - \kappa = 0.
\end{aligned}
\right.
\end{equation}
\end{small}%
It is obvious that the allocated bandwidth of each participating client $k$ is large than 0. Hence, we derive $\lambda_{2,k} = 0$ from the complementary relaxation condition $\lambda_{2,k} B_k^t = 0$. The KKT conditions, then, can be transformed into
\begin{small}
\begin{equation}
\left\{
\begin{aligned}
& \lambda_{4,k} \geq 0,\quad \sum_{k \in \mathcal K^t} B_k^t - B^{\max} = 0, \quad B_k^t > 0, \\
&  \textbf{In1:}\ \frac{\Gamma_k}{B_k^t \log_2(1 + \frac{p h_k^t}{B_k^t N_0})} \leq \tau^{\max} - \frac{D_k \Phi_k}{f}, \\
& \textbf{E1:}\ \lambda_{4,k} \big( \frac{\Gamma_k}{B_k^t \log_2(1 + \frac{p h_k^t}{B_k^t N_0})} + \frac{D_k \Phi_k}{f} -\tau^{\max} \big)= 0 , \\
& \textbf{E2:}\ (Q_k^t p \Gamma_k + \lambda_{4,k} \Gamma_k)\frac{ \frac{ph_k^t}{ph_k^t + B_k^tN_0} - \ln(1 + \frac{ph_k^t}{B_k^tN_0})}{ \frac{1}{\ln2}(B_k^t)^2 \ln^2 (1 + \frac{p h_k^t}{B_k^t N_0})} = \kappa.
\end{aligned}
\right. \label{equation: solve}
\end{equation}
\end{small}%
We observe that \textbf{In1} has a equivalent inequality $B_k^t \geq B_k^{t, \min}$, where $B_k^{t, \min}$ satisfies
\begin{equation}
\frac{\Gamma_k}{B_k^{t, \min} \log_2(1 + \frac{p h_k^t}{B_k^{t, \min} N_0})} = \tau^{\max} - \frac{D_k \Phi_k}{f}.
\label{equation: minimal bandwidth}
\end{equation}
Due to $\frac{\partial J_3}{\partial B_k^t} > 0$, the function in the left side of (\ref{equation: minimal bandwidth}), which is equal to the objective function in $\textbf{P4.2}'$ divided by $\frac{J_3}{Q_k^t p}$, monotonically increases with bandwidth increasing. Hence, we deduce that (\ref{equation: minimal bandwidth}) has a unique solution $B_k^{t,\min}$, and $B_k^t \geq B_k^{t, \min}$ is a necessary and sufficient condition of \textbf{In1}. As a transcendental equation, (\ref{equation: minimal bandwidth}) relies on numerical methods such as the Newton iteration method to solve. After obtaining $B_k^{t, \min}$, the feasibility of $\bm a^t$ can be judged by
\begin{equation}
\sum_{k \in \mathcal K^t} B_k^{t, \min} \leq B^{\max}.
\label{equation: bandwidth detect}
\end{equation}
\par
Unsatisfied (\ref{equation: bandwidth detect}) suggests that given $\bm a^t$ is infeasible. If (\ref{equation: bandwidth detect}) holds with equality, we can prove that the optimal bandwidth for $k\in \mathcal K^t$ is $B_k^{t*} = B_k^{t,\min}$. Otherwise, (\ref{equation: bandwidth detect}) is loose, and then we define $\kappa_k^t$ by
$
\kappa_k^t \triangleq Q_k^t p \Gamma_k \cdot \frac{ \frac{ph_k^t}{ph_k^t + B_k^{t, \min}N_0} - \ln(1 + \frac{ph_k^t}{B_k^{t, \min}N_0})}{ \frac{1}{\ln2}(B_k^{t, \min})^2 \ln^2 (1 + \frac{p h_k^t}{B_k^{t, \min} N_0})}
$
We can easily prove $\kappa_k^t < 0$. Without loss of generality, arrange $\kappa_k^t$ of participating clients in ascending order as $\kappa^t_{i_1} \leq \kappa^t_{i_2} \leq \kappa^t_{i_3} \leq \cdots \leq \kappa^t_{i_\varkappa} < 0$, where $\varkappa$ is the number of clients in $\mathcal K^t$. According to the function in the left side of \textbf{E2} monotonically increasing with $B_k^t$ and $\lambda_{4,k} > 0$, feasible $\kappa$ satisfies $\kappa \geq \kappa_{i_1}^t$. Consequently, the value of $\kappa$ is either between $\kappa_{i_j}^t$ and $\kappa_{i_{j+1}}^t$ or beyond $\kappa^t_{i_\varkappa}$, i.e.
\begin{equation}
\left\{
\begin{aligned}
& \kappa = \kappa_{i_1}; \\
& \kappa \in (\kappa_{i_j}, \kappa_{i_{j+1}}], \quad j=1,2,\cdots, \varkappa-1; \\
& \kappa \in (\kappa_{i_\varkappa}, + \infty).
\end{aligned}
\right.
\end{equation}
\par
If $\kappa = \kappa_{i_1}$, we can easily derive the same solution as strictly satisfied (\ref{equation: bandwidth detect}). And we directly take $\kappa \in (\kappa_{i_j}, \kappa_{i_{j+1}}]$ into consideration. For client $k = i_1, \cdots, i_j$, we can deduce $\lambda_{4, k} = 0$ and $B_k^t > B_k^{t, \min}$ from \textbf{E1} and \textbf{E2}. Similar to (\ref{equation: minimal bandwidth}), we firstly solve the boundary bandwidth $B_{k,{j+1}}^t$ satisfying
\begin{equation}
Q_k^t p \Gamma_k \frac{ \frac{ph_k^t}{ph_k^t + B_{k,{j+1}}^tN_0} - \ln(1 + \frac{ph_k^t}{B_{k,{j+1}}^t N_0})}{ \frac{1}{\ln2}(B_{k,{j+1}}^t)^2 \ln^2 (1 + \frac{p h_k^t}{B_{k,{j+1}}^t N_0})} = \kappa_{i_{j+1}}.
\label{equation: boundary bandwidth}
\end{equation}
Furthermore, due to $\lambda_{4,k} > 0$ for \textbf{E1} and \textbf{E2}, client $k = i_{j+1}, i_{j+2}, \cdots, i_\varkappa$ has $B_k^{t*} = B_k^{t, \min}$. As thus, a prerequisite inequality is formulated as
\begin{equation}
\textbf{Pre:} \sum_{k=i_1}^{i_j} B^t_{k,j+1} + \sum _{k=i_{j+1}}^{i_\varkappa} B_k^{t, \min} \geq B^{\max}.
\end{equation}
If \textbf{Pre} is not satisfied on $(\kappa_{i_j}, \kappa_{i_{j+1}}]$, the next interval $(\kappa_{i_{j+1}}, \kappa_{i_{j+2}}]$ will be considered till $(\kappa_{i_\varkappa}, +\infty)$. Without loss of generality, we suppose that \textbf{Pre} holds on $(\kappa_{i_j}, \kappa_{i_{j+1}}]$. Thus, the equation of $\kappa^*$ is formulated by
\begin{equation}
\sum_{k=i_1}^{i_j} B^{t*}_k(\kappa^*) + \sum _{k=i_{j+1}}^{i_\varkappa} B_k^{t, \min} = B^{\max}.
\label{eq4: bandwidth sum}
\end{equation}
Substituting $\lambda_{4, k} = 0$ into \textbf{E2}, the relation between $B^{t*}$  and $\kappa^*$ for client $k=i_1, \cdots, i_j$ is expressed by
\begin{equation}
Q_k^t p \Gamma_k \frac{ \frac{ph_k^t}{ph_k^t + B_k^{t*}N_0} - \ln(1 + \frac{ph_k^t}{B_k^{t*}N_0})}{ \frac{1}{\ln2}(B_k^{t*})^2 \ln^2 (1 + \frac{p h_k^t}{B_k^{t*} N_0})} = \kappa^*.
\label{equation: relation}
\end{equation}
Despite no closed-form function for $B_k^{t*}(\kappa^*)$, the Newton iteration method can be utilized since $\frac{d B_k^{t*}}{d\kappa^*}$ is equal to $\frac{1}{ \frac{d\kappa^*}{dB_k^{t*}}}$ according to implicit differentiation of (\ref{equation: relation}), so that $B_k^{t*}(\kappa^*)$ is solved for client $k=i_1, \cdots, i_j$.
\par
If there is no $(\kappa_{i_j}, \kappa_{i_{j+1}}]$ satisfying \textbf{Pre}, $(\kappa_{i_\varkappa}, + \infty)$ will finally be considered. And the corresponding formulation for client $k=i_1, i_2, \cdots, i_\varkappa$ is
\begin{equation}
\sum_{k=i_1}^{i_\varkappa} B^{t*}_k(\kappa^*)  = B^{\max}.
\label{eq4: bandwidth sum max}
\end{equation}
Similar to (\ref{eq4: bandwidth sum}), (\ref{eq4: bandwidth sum max}) can be solved by the Newton iteration method and implicit differentiation of (\ref{equation: relation}). To sum up, the optimal bandwidth for $k \in \mathcal K^t$ is (\ref{equation: optimal bandwidth}).
\begin{figure*}
\begin{equation}
B^{t*}_k =
\left\{
\begin{aligned}
& 0 \ {\rm for} \ k \in \mathcal K^t, & {\rm (\ref{equation: bandwidth detect})\ is\ unsatisfied}; \\
& B_k^{t, \min}\ {\rm for} \ k \in \mathcal K^t, & {\rm (\ref{equation: bandwidth detect})\ holds\ with\ equality}; \\
&  B_k^{t, \min}\ {\rm for} \ k=i_{j+1}, \cdots, i_{\varkappa};\ {\rm solve\, (\ref{eq4: bandwidth sum})\ and\ } \textbf{E2}\ {\rm for} \ k=i_1, \cdots, i_j, & {\rm (\ref{equation: bandwidth detect})\ is\ loose,\: } \textbf{Pre}\ {\rm is\ satisfied}; \\
& {\rm solve\ (\ref{eq4: bandwidth sum max})\ and\ } \textbf{E2}\ {\rm for} \ k \in \mathcal K^t, & {\rm (\ref{equation: bandwidth detect})\ is\ loose,\ but\ } \textbf{Pre}\ {\rm is\ not\ satisfied}.
\end{aligned}
\right. \label{equation: optimal bandwidth}
\end{equation}
\hrulefill
\end{figure*}
\begin{algorithm}[t]
  \caption{Immune Algorithm for Client Scheduling}\label{algorithm: immune algorithm}
  \KwIn{\textcolor{black}{$S = 20, G = 10, \mu = 5, z = 0.175$}}
  \KwOut{optimal participation vector\ $\bm a^{t*}$}
  Initialize the set of $S$ antibodies randomly as $\mathcal A^0 = \{ \bm a_1^{t, 0}, \bm a_2^{t, 0}, \cdots, \bm a_S^{t,0} \}$\;
  \For{$g = 0, 1, \cdots, G-1$}
  {
    \For{$s = 1, 2, \cdots, S$}
    {
        Compute ${\rm aff}({\bm a_s^{t, g}})$ and ${\rm con}(\bm a_s^{t, g})$ for antibody $\bm a_s^{t, g}$ according to (\ref{equation: affinity}) and (\ref{equation: concentration})\;
        Compute ${\rm inc}({\bm a_s^{t, g}})$ for antibody $\bm a_s^{t,g}$ according to (\ref{equation: incentive})\;
    }
    Select the top $\frac{S}{\mu}$ antibodies with the highest incentive values from $\mathcal A^g$ to constitute $\mathcal A_{\rm imm}^{g+1}$ \;
    Clone each antibody of $\mathcal A_{\rm imm}^{g+1}$ $\mu$ times to obtain the clone set $\mathcal A_{\rm clo}^{g+1}$\;
    Mutate all antibodies in the clone set $\mathcal A_{\rm clo}^{g+1}$ with the mutation rate $z$ of each gene to generate the mutation set $\mathcal A_{\rm mut}^{g+1}$\;
    Compute the affinity values of all antibodies in $\mathcal A_{\rm mut}^{g+1}$\;
    Select top $\frac{\mu S - S}{\mu}$ antibodies with the highest affinity values from $\mathcal A_{\rm mut}^{g+1} \bigcup \mathcal A_{\rm imm}^{g+1}$\;
    Obtain the next antibody set $\mathcal A^{g+1}$ with the selected top $\frac{\mu S - S}{\mu}$ antibodies and the other $\frac{S}{\mu}$ randomly generated antibodies\;
  }
  \Return $\bm a^{t*} = \arg\max_{a_s^{t, G}\in \mathcal A^G} {\rm aff}(a_s^{t, G})$\;
\end{algorithm}
\subsection{Combinatorial Subproblem}
Although $\textbf{P4.2}'$ for attains the optimal bandwidth vector for any given $\bm a^t$, a closed-form solution remains intractable, leading to implicit $J_2(\bm a^t)$ in $\textbf{P4.2}'$. In such cases, the combinatorial optimization problem \textbf{P4.1} can rely on heuristic algorithms to search for optimal $\bm a^t$. In wireless channel allocation \cite{immune2} and job scheduling \cite{immune1}, the immune algorithm has demonstrated superior performance. Furthermore, for $J_1(\bm a^t, \bm B^{t*})$ in \textbf{P4.1}, $\bm a^t$ is firmly related to which modalities participate in training in the $t$-th communication round. Various participation of modalities causes significant variations in $J_1(\bm a^t, \bm B^{t*})$. And the immune algorithm ensures antibody diversity within a generation through antibody concentration \cite{immune3}, which is suitable for client scheduling involving various participation of modalities.
\par
Set the generation number $g:=0$ and randomly generate the initial antibody set $\mathcal A^0 = \{ \bm a_1^{t,0}, \bm a_2^{t, 0}, \cdots , \bm a_S^{t, 0} \}$. According to $J_2(\bm a^t)$, the affinity function of antibody $\bm a_s^{t, g}, s=1, 2, \cdots, S$ is formulated as
\begin{equation}
{\rm aff}(\bm a_s^{t,g}) = (J_2^{\max}, J_2(\bm a_s^{t, g}))^\iota,
\label{equation: affinity}
\end{equation}
where $J_2^{\max}\triangleq \max_{s=1, 2, \cdots, S} J_2(\bm a_s^{t, g})$ and $\iota$ is a exponential coefficient to adjust dispersion of the affinity function. As for infeasible $\bm a_s^{t, g}$, its affinity function is set to 0.
\par
Computing the similarity from $\bm a_s^{t, g}$ to each antibody in $\mathcal A^g$, the antibody concentration is
\begin{equation}
{\rm con} (\bm a_s^{t,g}) = \frac{1}{S} \sum_{i = 1}^S {\rm sim} (\bm a_s^{t,g}, \bm a_i^{t, g}),
\label{equation: concentration}
\end{equation}
where the similarity function ${\rm sim}(\cdot)$ is determined by comparing the Hamming distance ${\rm dis}(\cdot)$ to a threshold ${\rm Dis}$, i.e.,
\begin{equation}
{\rm sim} (\bm a_s^{t,g}, \bm a_i^{t, g}) =
\begin{cases}
1, & {\rm dis} (\bm a_s^{t,g}, \bm a_i^{t, g}) \leq {\rm Dis}; \\
0, & {\rm dis} (\bm a_s^{t,g}, \bm a_i^{t, g}) > {\rm Dis}.
\end{cases}
\end{equation}
\par
The incentive function balancing the affinity function and the diversity function is
\begin{equation}
{\rm inc} (\bm a_s^{t,g}) = \epsilon_1 {\rm aff} (\bm a_s^{t,g}) - \epsilon_2 {\rm con} (\bm a_s^{t,g}).
\label{equation: incentive}
\end{equation}
According to incentive values of $\mathcal A^g$, the top $\frac{S}{\mu}$ antibodies are selected to constitute $\mathcal A_{\rm imm}^{g+1}$ for $\mu$-fold cloning and mutation. Select the top $\frac{\mu S - S}{\mu}$ antibodies from $\mathcal A_{\rm imm}^{g+1}$ and mutant antibodies, and generate randomly $\frac{S}{\mu}$ antibodies. The next antibody set $\mathcal A^{g+1}$ is composed by the selected $\frac{\mu S - S}{\mu}$ antibodies and the generated $\frac{S}{\mu}$ antibodies. Similarly, we can obtain $\mathcal A^{g+2}, A^{g+3}, \cdots$ till the maximal generation number $G$. And $\bm a^{t*}$ is the antibody with the maximal affinity in $\mathcal A^G$. The detailed process is presented in \textbf{Algorithm \ref{algorithm: immune algorithm}}.
\subsection{Computational Complexity Analysis}
The complexity of solving the continuous subproblem comes from the Newton iteration in (\ref{equation: minimal bandwidth}), (\ref{equation: boundary bandwidth}), and (\ref{eq4: bandwidth sum}) or (\ref{eq4: bandwidth sum max}). Given convergence thresholds $\epsilon_\tau, \epsilon_B, \epsilon_\kappa$ for $\tau^{\max}$, $B^{\max}$ and $\kappa$, the computational complexity of the worst case is $\mathcal O\big(U^2 \log(\frac{1}{\epsilon_\kappa}) + U\log(\frac{1}{\epsilon_B}) \log(\frac{1}{\epsilon_\kappa}) + U\log (\frac{1}{\epsilon_\tau}) \big)$.
As for the combinatorial subproblem, the complexity is positively related to $S$ and $G$. Thus the computational complexity of \textbf{Algorithm \ref{algorithm: immune algorithm}} is $\mathcal O(SG)$.
The total complexity of the solution is $\mathcal O\big( SG  U^2 \log(\frac{1}{\epsilon_\kappa}) + SG U\log(\frac{1}{\epsilon_B}) \log(\frac{1}{\epsilon_\kappa}) + SG U\log (\frac{1}{\epsilon_\tau}) \big)$. However, benefiting from the parallel computing among antibodies and clients, the actual computational time can be reduced to $\frac{1}{SU}$ of the computational complexity.
\section{Numerical Results}
In the experiment, a base station (BS) connected to a server and $K$ clients are set in a circular network area to collaboratively complete wireless MFL tasks. All clients are uniformly distributed within a 500-meter radius. Other parameters of wireless MFL are listed in Table \ref{tab: parameter}.
\begin{table}[ht]
  \centering
  \caption{System Parameters}
  \begin{tabular}{c|c|c|c}
    \hline \hline
    \textbf{Parameter} & \textbf{Value} & \textbf{Parameter} & \textbf{Value} \\ \hline
    $B^{\max}$ & 10 MHz & $\tau^{\max}$ & 0.01 s \\ \hline
    $p$ & 23 dBm & $N_0$ & -174 dBm/Hz \\ \hline
    $\ell_{\rm audio}^{\rm CREMA-D}$ & 562400 & $\ell_{\rm image}^{\rm CREMA-D}$ &  557056 \\ \hline
    $\ell_{\rm audio}^{\rm IEMOCAP}$ & 562400 & $\ell_{\rm text}^{\rm IEMOCAP}$ &  1145280 \\ \hline
    $\beta_{\rm audio}^{\rm CREMA-D}$ & 2000 & $\beta_{\rm photo}^{\rm CREMA-D}$ & 8000 \\ \hline
    $\beta_{\rm audio}^{\rm IEMOCAP}$ & 2000 & $\beta_{\rm text}^{\rm IEMOCAP}$ & 4500 \\ \hline
    $K$ & 10 & $E^{\rm add}$ & 0.01 J \\ \hline
    $f$ & $1.55\times10^9$ Hz & $\alpha$ & $10^{-27}$ \\ \hline
    \hline
  \end{tabular}
  \label{tab: parameter}
\end{table}
\par
\textbf{Datasets.} We utilize two multimodal datasets to validate our JCSBA algorithm. One multimodal dataset is Crowd-Sourced Emotional Multimodal Actors Dataset \cite{CREMAD} (CREMA-D). Such an audio-visual dataset consists of 7442 video clips, which are labeled by one of six emotion states (happy, sad, angry, fear, disgust, and neutral). An image frame in the middle of each video clip is an image sample. A total audio clip of each video clip is an audio sample. Both of them constitute a multimodal sample. The other multimodal dataset is Interactive Emotional Dyadic Motion Capture DataBase \cite{iemocap} (IEMOCAP). Such a multimodal dataset consists of 10039-turn dialogue, which is labeled by one of ten emotion states (happy, sad, anger, surprise, neutral, fear, disgust, frustrated, excited, other). The subtitle of each turn is a text sample. A total audio clip of each turn is an audio sample. Similarly, both of them constitute a multimodal sample. \textcolor{black}{Modality heterogeneity is quantified by a missing modality ratio $\omega$\cite{distribution}. Specifically, we set $\omega_{\rm audio}^{\rm CREMA-D} = \omega_{\rm photo}^{\rm CREMA-D} = \omega_{\rm audio}^{\rm IEMOCAP} = \omega_{\rm text}^{\rm IEMOCAP} = 0.3$, where $\omega_{\rm audio}^{\rm CREMA-D} = 0.3$ indicates that $30\%$ of clients lack the audio modality, and others in turn.}
\par
\textbf{Models.} The multimodal model for CREMA-D consists of an audio submodel and an image submodel. The audio submodel is composed of a unidirectional two-layer Long Short-Term Memory (LSTM), a 50-neuron hidden layer, and a 6-neuron output layer. As for the detailed architecture of the audio LSTM, the dropout rate is 0.1, the input size of audio features is 11, and both the hidden size and the output size are 50. The image submodel is a convolutional neural network (CNN) composed of 3 convolutional layers with the 3-stride $5\times 5$ maxpooling, 2 hidden layers, and an output layer with 6 neurons. The three convolutional layers have 16 kernels with $3\times 5 \times 5$, $16 \times 5 \times 5$ and $16 \times 5 \times 5$ shapes, respectively. And the two hidden layers have 64 and 32 neurons, respectively. The multimodal model for IEMOCAP consists of an audio submodel and a text submodel. The audio submodel is the same as that for CREMA-D, except for the 10-neuron output layer. The text submodel has a unidirectional two-layer LSTM, a 60-neuron hidden layer, and a 10-neuron output layer. In the text LSTM, the dropout rate is 0.1, the input size of text features is 100, and both the hidden size and the output size are 60.
\par
\textbf{Baselines.} To compare MFL performance, 2 conventional algorithms serve as baselines: (a) the algorithm with scheduling clients and allocating bandwidth randomly (see the line labeled with ``Random''), (b) the round-robin scheduling algorithm with allocating bandwidth equally (``Round-Robin''). Moreover, we simulate other 2 MFL algorithms: (c) the algorithm in \cite{select_client} which utilizes fixed selection ratios for clients with different combinations of modalities, and then selects them from clients with each combination of modalities according to the distances of their local models to the initial model (``Selection''), (d) the algorithm in \cite{dropout} which employs modality dropout for multimodal clients with a probability (``Dropout''). All curves of algorithms are obtained with the average of 10 experimental results.
\par
\textcolor{black}{\textbf{Runtime.} The maximal runtime of our problem solution in Section V is 0.008 s in each communication round, which reduces more than $90\%$ of 0.097 s required by the simulated annealing algorithm. The efficiency of our solution is due to the parallel computing among antibodies and considering the antibody concentration to search the optimal point. Since 0.008 s $< \tau^{\max}$, the server can obtain the optimal point in time for client scheduling and bandwidth allocation in the next communication round. All results are obtained with Python, using the Siyuan-1 cluster with 2 x Intel Xeon ICX Platinum 8358.}
\par
In Section VI-A, the total energy consumption and the final performance of JCSBA with different $V$ values are given to select $V$. In Section VI-B and Section VI-C, energy consumption, multimodal performance, and unimodal performance of 5 algorithms are compared based on CREMA-D and IEMOCAP, respectively.
\begin{figure}[t]
  \centering
  \subfigure[CREMAD]{\includegraphics[width=0.24\textwidth]{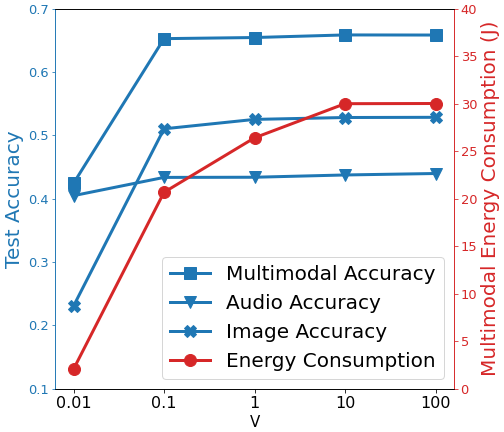}}
  \subfigure[IEMOCAP]{\includegraphics[width=0.24\textwidth]{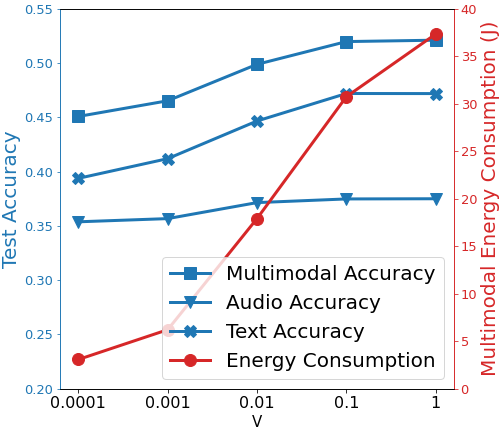}}
  \caption{Total energy consumption, final multimodal and unimodal performance of JCSBA with different $V$ values.}
  \label{fig: trade-off}
  \vspace{-3mm}
\end{figure}

\subsection{Trade-off between Performance and Energy Consumption}
Fig. \ref{fig: trade-off} shows the impact of different $V$ values on energy consumption, multimodal performance, and unimodal performance. From Fig. \ref{fig: trade-off}(a), it can be seen that the energy consumption rises, and the multimodal accuracy and the two unimodal accuracies decrease with $V$ rising. This is due to the fact that a large $V$ in (\ref{equation: Lyapunov problem}) emphasizes the impact of performance and neglects energy consumption. Consequently, the JCSBA algorithm schedules more clients to pursue better performance at the cost of more energy consumption. When $V=0.01$, the image accuracy is rarely low, whereas the multimodal and audio accuracies are relatively higher. This is because the computation on the image CNN is more complex than the audio LSTM, resulting in the former consuming much more energy than the latter. Hence, with rather limited energy, the JCSBA algorithm prioritizes updating the audio submodel to increase the multimodal performance. The slight gap between the audio accuracy and the image accuracy indicates that the multimodal model mainly relies on the audio submodel. Increasing $V$ from 0.01 to 1, the image submodel has enough energy to update, which then increases the image accuracy and the multimodal accuracy. However, all three accuracy curves and the energy curve get flat for $V > 1$, demonstrating that the energy allocated to MFL is used up. With a trade-off between the MFL performance and the energy consumption, we choose $V=1$ for subsequent experiments in Section VI-B.
\par
As shown in Figure \ref{fig: trade-off}(b), the degradations of both multimodal and unimodal performance degradation on $V=0.0001$ are not particularly obvious. The underlying reason is that the audio LSTM and the text LSTM are in the energy consumption of computation. Therefore, the JCSBA algorithm still updates two submodels under the low-energy demand. Benefiting from the fast convergence of the audio modality, the audio modality with low energy and fewer participating clients exhibits less noticeable decline of the performance than the text modality. For $V\geq0.1$, three accuracy curves get flat despite a rising energy curve, indicating that the multimodal model has reached its optimal performance within the current architecture. Therefore, $V=0.1$ is selected for subsequent experiments in Section VII-C.
\subsection{MFL on CREMA-D}
\begin{figure}[t]
  \centering
  \subfigure[Multimodal Accuracy]{\includegraphics[width=0.24\textwidth]{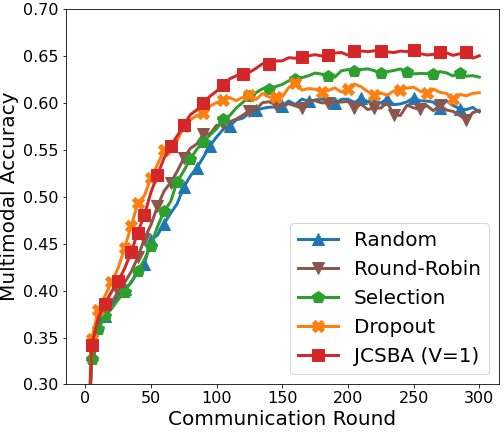}}
  \subfigure[Energy Consumption]{\includegraphics[width=0.24\textwidth]{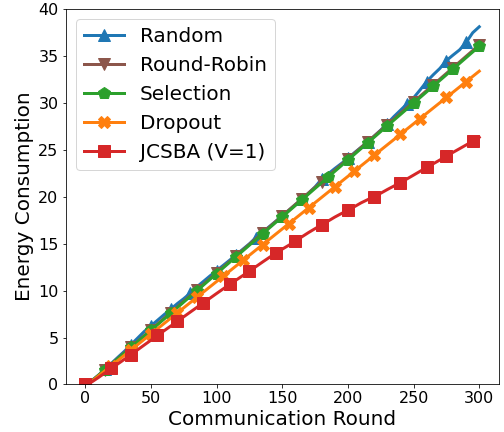}}
  \\
  \subfigure[Audio Accuracy]{\includegraphics[width=0.24\textwidth]{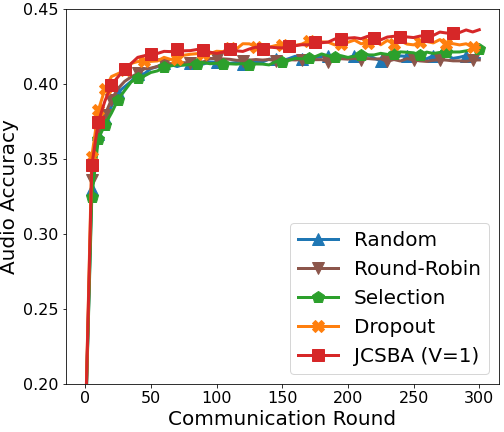}}
  \subfigure[Image Accuracy]{\includegraphics[width=0.24\textwidth]{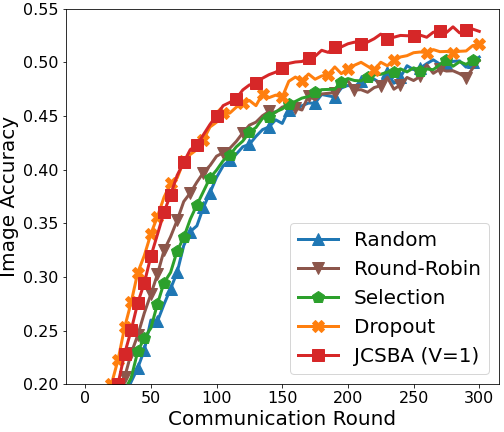}}
  \caption{Multimodal accuracy, unimodal accuracy and energy consumption curves in all communication rounds on CREMA-D.}
  \label{fig: human emotion}
  \vspace{-3mm}
\end{figure}

\begin{table*}[t]
\centering
\caption{Multimodal accuracies and unimodal accuracies on CREMA-D and IEMOCAP}
\begin{tabular}{c|c||ccccc}
\hline\hline
Dataset & Algorithm  & Random  & Round Robin & Selection & Dropout & JCSBA            \\ \hline
\multirow{3}{*}{CREMA-D}    & Multimodal & 61.93\% & 61.37\%     & 64.33\%   & 62.65\% & \textbf{65.99\%} \\
                            & Photo      & 50.89\% & 50.95\%     & 51.03\%   & 52.33\% & \textbf{53.68\%} \\
                            & Audio      & 42.26\% & 42.04\%     & 42.47\%   & 43.37\% & \textbf{43.90\%} \\ \hline
\multirow{3}{*}{IEMOCAP}    & Multimodal & 49.66\% & 49.89\%     & 50.97\%   & 50.81\% & \textbf{52.32\%} \\
                            & Text       & 45.38\% & 45.66\%     & 46.03\%   & 46.88\% & \textbf{47.69\%} \\
                            & Audio      & 36.96\% & 37.05\%     & 37.01\%   & 37.86\% & \textbf{38.26\%} \\ \hline\hline
\end{tabular}
\label{tab: accuracy}
\end{table*}
In Fig. \ref{fig: human emotion}, we compare multimodal performance, energy consumption, audio performance, and image performance of five algorithms. It is evident that the JCSBA algorithm achieves the fastest convergence and the highest accuracy in both the multimodal accuracy and the two unimodal accuracies while consuming the least energy. From Fig. \ref{fig: human emotion}(a), we can observe that the Selection algorithm performs better than the Dropout algorithm. It is attributed to the fact that the Dropout algorithm degrades multimodal datasets into unimodal datasets with a probability. As thus, there are fewer updates for the multimodal model compared to the Selection algorithm, which schedules clients with multimodal datasets without degradation. In Fig. \ref{fig: human emotion}(c) and (d), conversely, the Dropout algorithm exhibits higher unimodal accuracies than the Selection algorithm on account of the modal dropout. However, its unimodal accuracies remain inferior to the unimodal accuracies of the JCSBA algorithm. This is because the JCSBA algorithm adds the unimodal loss in (\ref{equation: local loss}) for clients with multimodal datasets, equivalent to the local loss on the degraded multimodal datasets. And the existing multimodal loss in (\ref{equation: local loss}) avoids the fewer updates of the Dropout algorithm on multimodal datasets. Furthermore, the JCSBA algorithm optimizes client scheduling and bandwidth allocation under latency and bandwidth constraints, leading to more efficient energy utilization, as demonstrated in Fig. \ref{fig: human emotion}(b). In contrast, the Random algorithm and the Round-Robin algorithm lack wise scheduling and optimization strategies, such that the two conventional algorithms perform terribly and consume much energy. The detailed multimodal and unimodal accuracies of all algorithms are summarized in Table \ref{tab: accuracy}.
\par
Comparing Fig. \ref{fig: human emotion}(c) and (d), we observe that the audio submodel converges faster than the image submodel. The different convergence speeds of the two modalities account for this phenomenon. As we can imagine, after the 50th communication round, the Dropout algorithm still degrades multimodal datasets into audio datasets, and the Selection algorithm maintains a fixed ratio of scheduling clients with audio datasets. Such two methods not only enhance the audio accuracy a little, but also consume energy in vain. In contrast, the JCSBA algorithm detects the modality imbalance by means of (\ref{equation: theorem1}) in \textbf{Theorem \ref{theorem1}}, and prioritizes scheduling clients with image data, aiming at the unconverged modality.
\begin{figure}[t]
  \centering
  \subfigure[Multimodal Accuracy]{\includegraphics[width=0.24\textwidth]{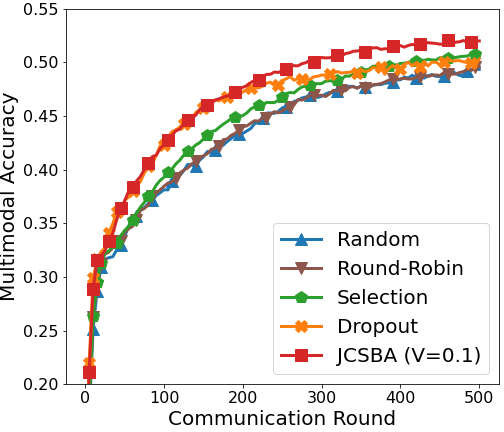}}
  \subfigure[Energy Consumption]{\includegraphics[width=0.24\textwidth]{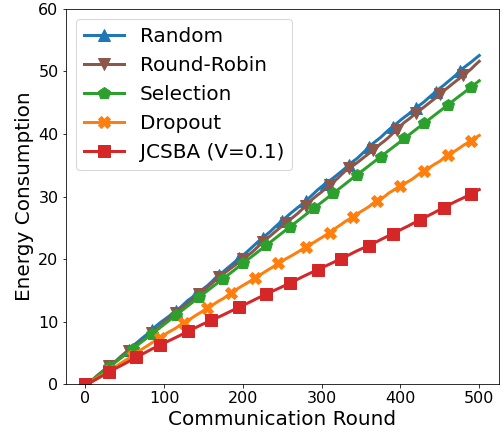}}
  \\
  \subfigure[Audio Accuracy]{\includegraphics[width=0.24\textwidth]{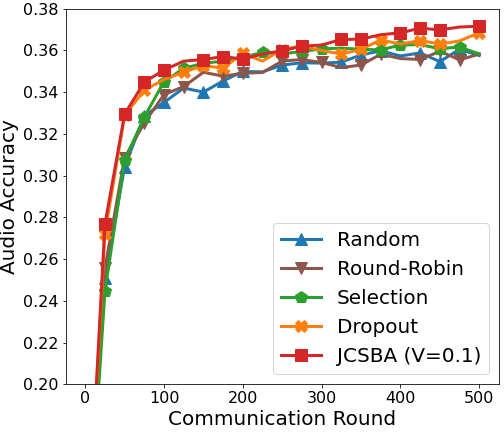}}
  \subfigure[Text Accuracy]{\includegraphics[width=0.24\textwidth]{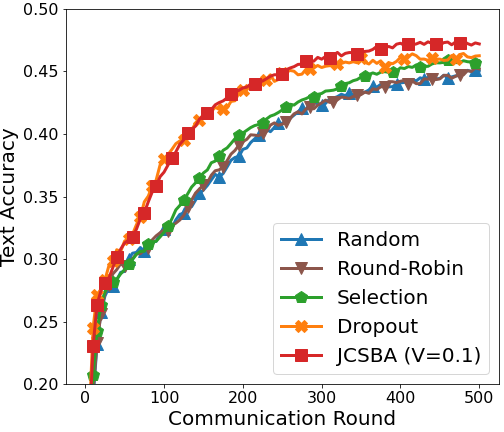}}
  \caption{Multimodal accuracy, unimodal accuracy and energy consumption curves in all communication rounds on IEMOCAP.}
  \label{fig: speech emotion}
  \vspace{-3mm}
\end{figure}
\vspace{-5mm}
\subsection{MFL on IEMOCAP}
Fig. \ref{fig: speech emotion} shows multimodal performance, unimodal performance, and energy consumption of five algorithms on IEMOCAP. From Fig. \ref{fig: speech emotion} (a)-(d), we observe the following trends: The JCSBA algorithm consistently outperforms other algorithms in the multimodal performance, unimodal performance, and energy consumption. The Dropout algorithm converges faster than the Selection algorithm in the unimodal accuracy on the text modality. However, despite lagging behind, the Selection algorithm eventually surpasses the Dropout algorithm in the multimodal accuracy. Moreover, the Random algorithm and Round-Robin algorithms exhibit the worst performance and highest energy consumption. The above results align with the results in Section VI-B and have similar reasons, validating the general applicability of the JCSBA algorithm. The detailed multimodal and unimodal accuracies of all algorithms are summarized in Table \ref{tab: accuracy}.
\par
Comparing Fig. \ref{fig: speech emotion} (c) and (d), we can see that the audio modality converges faster than the text modality on IEMOCAP. This further demonstrates that different convergence rates among modalities are common in multimodal learning. Hence, it is vital for an MFL algorithm to schedule clients with modality heterogeneity in consideration of the convergence process of all modalities.

\section{Conclusions}
In this paper, we have proposed the JCSBA algorithm for wireless MFL on modal heterogeneity to enhance both the multimodal performance and the unimodal performance. The wireless MFL architecture has employed the decision-level fusion and added the unimodal loss in local updates and the objective function. Building on the architecture, the JCSBA algorithm has adaptively scheduled clients for convergence of all modalities and allocated bandwidth to ensure clients with modal heterogeneity participating in MFL within latency. The experiments on multimodal datasets have validated the performance gain of our JCSBA algorithm.
\par
As for potential future works, modality heterogeneity in MFL could be extended to the sample level, where different samples of a client correspond to heterogeneous modalities \cite{contrast_representation}. Additionally, clients can flexibly select a subset of their available modalities for training according to wireless constraints and MFL performance. In such cases, modeling the latency and the energy consumption would become quite difficult, making performance analysis and optimization even more challenging but also more meaningful.
\appendix
\subsection{Proof of Theorem 1 \label{proof}}
With \textbf{Assumption \ref{assumption: Lipschitz}}, the difference of loss functions are transformed into the difference of models as
\begin{equation}
H(\bm\theta^t) - H(\bm\psi^t) \leq | H(\bm\theta^t) - H(\bm\psi^t) | \leq \rho \| \bm\theta^t - \bm\psi^t \|.
\label{proof: theorem1_1}
\end{equation}
Each model can be divided into $M$ submodels, which helps to express the norm as
\begin{equation}
\rho \| \bm\theta^t - \bm\psi^t \| = \rho \sqrt{\sum_{m=1}^M \| \bm\theta^t_m - \bm\psi^t_m \|^2}.
\label{proof: theorem1_2}
\end{equation}
\par
For modality $m$, there is either $m \in \mathcal M^t$ or $m \notin \mathcal M^t$. With $\bm\theta^t_{{\rm g}, m} = \bm\theta^{t-1}_{{\rm g}, m} $ for $m \notin \mathcal M^t$ and \textbf{Definition \ref{definition: auxiliary parameter}}, the difference of the two unimodal submodels for $m \notin \mathcal M^t$ is
\begin{equation}
\| \bm\theta_{{\rm g}, m}^t - \bm\psi_m^t \|^2 = \eta^2 \| \nabla H(\bm\theta_{{\rm g}, m}^{t-1}) \|^2 \leq \eta^2 (\zeta_m^{t-1})^2.
\label{proof: theorem1_3}
\end{equation}
As for $m\in \mathcal M^t$, (\ref{equation: participation local update}) and \textbf{Definition \ref{definition: auxiliary parameter}} are used to expand the difference as
\begin{align}
\| \bm\theta_{{\rm g}, m}^t - \bm\psi^t_m \|^2 = & \eta^2\Big\| \sum_{k \in \mathcal K_m^t} (w_{k,m}^t - \bar w_{k,m})\nabla H_k(\bm\theta_{{\rm g}, m}^{t-1}) \notag \\
& - \!\!\! \sum_{k \in \mathcal K_m \backslash \mathcal K_m^t}\!\!\! \bar w_{k,m} \nabla H_k(\bm\theta_{{\rm g}, m}^{t-1}) \Big\|^2. \label{proof: theorem1_3.1}
\end{align}
Adding two zero terms $ \big[ \sum_{k \in \mathcal K_m^t} (w_{k,m}^t - \bar w_{k,m}) \nabla H (\bm\theta_{{\rm g}, m}^{t-1}) $
$ - \sum_{k \in \mathcal K_m^t} (w_{k,m}^t  - \bar w_{k,m})  \nabla H (\bm\theta_{{\rm g}, m}^{t-1}) \big] $ and $\big[ \nabla H(\bm \theta^{t-1}_{ {\rm g}, m} )$ $-\sum_{k \in \mathcal K_m^t} \bar w_{k,m} \nabla H (\bm\theta_{{\rm g}, m}^{t-1})\big]$  into (\ref{proof: theorem1_3.1}), we then rearrange and combine terms to obtain an equation as
\begin{align}
& \| \bm\theta_{{\rm g}, m}^t - \bm\psi^t_m \|^2 \label{proof: theorem1_4.1} \\
& = \eta^2 \Big\| \sum_{k \in \mathcal K_m^t} (w_{k,m}^t - \bar w_{k,m}) \nabla H_k(\bm\theta_{{\rm g}, m}^{t-1}) \notag \\
& \qquad\quad - \sum_{k \in \mathcal K_m^t} (w_{k,m}^t - \bar w_{k,m}) \nabla H(\bm\theta_{{\rm g}, m}^{t-1}) \notag \\
& \qquad \quad + \!\!\!\!\!\! \sum_{k \in \mathcal K_m \backslash \mathcal K_m^t} \!\!\!\!\!\! \bar w_{k,m} \nabla H(\bm\theta_{{\rm g}, m}^{t-1}) - \!\!\!\!\!\! \sum_{k \in \mathcal K_m \backslash \mathcal K_m^t} \!\!\!\!\!\! \bar w_{k,m} \nabla H_k(\bm\theta_{{\rm g}, m}^{t-1}) \Big\|^2 \notag \\
& = \eta^2 \Big\| \sum_{k \in \mathcal K_m^t} \! (w_{k,m}^t - \bar w_{k,m}) (\nabla H_k(\bm\theta_{{\rm g}, m}^{t-1}) - \nabla H(\bm\theta_{{\rm g}, m}^{t-1})) \notag \\
& \qquad \quad- \!\!\!\!\!\! \sum_{k \in \mathcal K_m \backslash \mathcal K_m^t} \!\!\!\!\!\! \bar w_{k,m} (\nabla H_k(\bm\theta_{{\rm g}, m}^{t-1})\! - \! \nabla H(\bm\theta_{{\rm g}, m}^{t-1})) \Big\|^2 \notag.
\end{align}
Dividing the square norm in (\ref{proof: theorem1_4.1}) into two terms according to $\| \bm \varphi_1 - \bm\varphi_2 \|^2 \leq \| \bm\varphi_1 \|^2 + 2 \| \bm\varphi_2 \|^2$, we then obtain
\begin{align}
& \| \bm\theta^t_{{\rm {g}, m}} - \bm\psi^t_m \|^2 \label{proof: theorem1_5}  \\
& \leq 2\eta^2 \Big\| \sum_{k \in \mathcal K_m^t} (w_{k,m}^t - \bar w_{k,m}) (\nabla H_k(\bm\theta_{{\rm g}, m}^{t-1}) - \nabla H(\bm\theta_{{\rm g}, m}^{t-1}) \Big\|^2 \notag \\
& \quad + 2\eta^2 \Big\| \sum_{k \in \mathcal K_m \backslash \mathcal K_m^t} \bar w_{k,m} (\nabla H_k(\bm\theta_{{\rm g}, m}^{t-1}) - \nabla H(\bm\theta_{{\rm g}, m}^{t-1}) \Big\|^2. \notag
\end{align}
The weighted sum of gradient differences in (\ref{proof: theorem1_5}) is hard to tackle. To construct $\| \nabla H_k(\bm\theta_m^{t-1}) - \nabla H(\bm\theta^{t-1}) \|$ with Jensen's inequality, we first normalize the weights as
\begin{align}
& \| \bm\theta^t_{{\rm g}, m} - \bm\psi^t_m \|^2 \label{proof: theorem1_6} \\
& \leq \Big\| \!\! \sum_{k \in \mathcal K_m^t} \!\!\! \frac{w_{k,m}^t - \bar w_{k,m}}{1 - \sum_{i \in \mathcal K_m^t} \! \bar w_{i,m}} (\nabla H_k(\bm\theta_{{\rm g},m}^{t-1}) - \nabla H(\bm\theta_{{\rm g},m}^{t-1})) \Big\|^2 \notag \\
& \quad \times 2\eta^2(1 - \sum_{k \in \mathcal K_m^t} \bar w_{k,m})^2 + 2\eta^2(1 - \sum_{k \in \mathcal K_m^t} \bar w_{k,m})^2 \times \notag \\
& \quad \Big\| \!\! \sum_{k \in \mathcal K_m \backslash \mathcal K_m^t} \!\!\!\! \frac{\bar w_{k,m}}{1 - \sum_{i \in \mathcal K_m^t} \bar w_{i,m}} (\nabla H_k(\bm\theta_{{\rm g},m}^{t-1}) - \nabla H(\bm\theta_{{\rm g},m}^{t-1})) \Big\|^2 \notag
\end{align}
Since $\| \cdot \|^2$ is convex, Jensen' inequality can be applied in (\ref{proof: theorem1_6}). And \textbf{Assumption \ref{assumption: divergence}} is then used to enlarge the gradient difference into
\begin{align}
& \| \bm\theta^t_{{\rm g}, m} - \bm\psi^t_m \|^2 \label{proof: theorem1_7} \\
& \leq \sum_{k \in \mathcal K_m^t} (w_{k,m}^t - \bar w_{k,m}) \| \nabla H_k(\bm\theta_{{\rm g},m}^{t-1}) - \nabla H(\bm\theta_{{\rm g},m}^{t-1}) \|^2 \notag \\
& \quad \times 2 \eta^2 (1 - \sum_{k \in \mathcal K_m^t} \bar w_{k,m}) + 2 \eta^2 (1 - \sum_{k \in \mathcal K_m^t} \bar w_{k,m}) \times \notag \\
& \quad \!\!\! \sum_{k \in \mathcal K_m \backslash \mathcal K_m^t} \!\!\! \bar w_{k,m} \| \nabla H_k(\bm\theta_{{\rm g},m}^{t-1}) - \nabla H(\bm\theta_{{\rm g},m}^{t-1})\|^2 \notag
\end{align}
\begin{align}
& \leq 2 \eta^2 \Big( \sum_{k \in \mathcal K_m} (w_{k,m}^t - a_k^t \bar w_{k,m}) \! + \!\!\! \sum_{k \in \mathcal K_m} (1-a_k^t) \bar w_{k,m} \Big) \notag \\
& \quad \Big(1 - \!\!\! \sum_{k \in \mathcal K_m^t} \bar w_{k,m}\Big) \| \nabla H_k(\bm\theta_{{\rm g},m}^{t-1}) - \nabla H(\bm\theta_{{\rm g},m}^{t-1}) \|^2 \notag \\
& \leq 2(1 - \!\!\!\! \sum_{k \in \mathcal K_m} \!\!\! a_k^t \bar w_{k,m}) \!\!\! \sum_{k \in \mathcal K_m} \!\! (w_{k,m}^t \!\! + \bar w_{k,m} \!\! - 2a_k^t \bar w_{k,m}) (\delta_{k,m}^{t-1})^2. \notag
\end{align}
\par
Finally, (\ref{proof: theorem1_3}) and (\ref{proof: theorem1_7}) are summed up over $m \notin \mathcal M^t$ and $m \in \mathcal M^t$, respectively, and then the sums are substituted into (\ref{proof: theorem1_2}) to prove \textbf{Theorem \ref{theorem1}}.

\bibliographystyle{IEEEtran}
\bibliography{summary}

\begin{IEEEbiography}[{\includegraphics[width=1in,height=1.25in,clip,keepaspectratio]{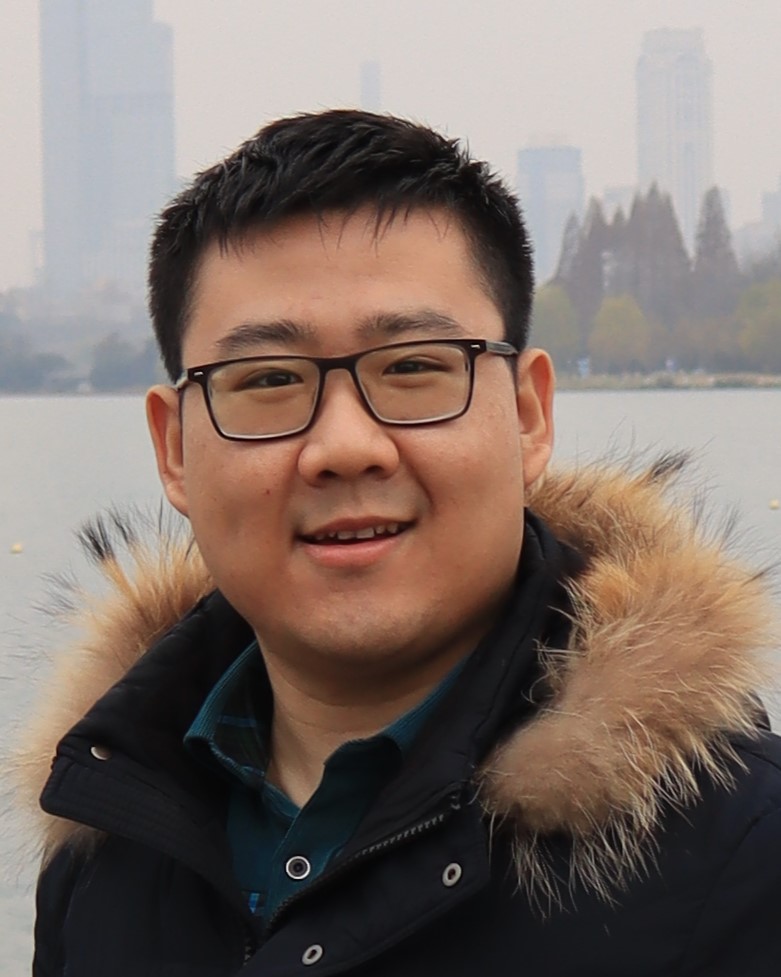}}]
{Xuefeng Han}~received the B.S. degree in communication engineering from University of Electronic Science and Technology of China in 2020, and received the Ph.D. degree from Department of Electronic Engineering, Shanghai Jiao Tong University (SJTU) in 2025. His research interests include federated learning, lightweight neural network, multimodal learning and resource management in future wireless networks.
\end{IEEEbiography}

\begin{IEEEbiography}[{\includegraphics[width=1in,height=1.25in,clip,keepaspectratio]{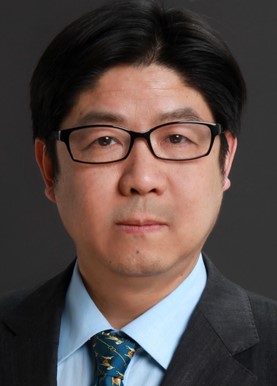}}]
{Wen Chen}~(M’03–SM’11)~received BS and MS from Wuhan University, China in 1990 and 1993 respectively, and PhD from University of Electro-communications, Japan in 1999. He is now a tenured Professor with the Department of Electronic Engineering, Shanghai Jiao Tong University, China. He is a fellow of Chinese Institute of Electronics and the distinguished lecturers of IEEE Communications Society and IEEE Vehicular Technology Society. He also received Shanghai Natural Science Award in 2022. He is the Shanghai Chapter Chair of IEEE Vehicular Technology Society, a vice president of Shanghai Institute of Electronics, Editors of IEEE Transactions on Wireless Communications, IEEE Transactions on Communications, IEEE Access and IEEE Open Journal of Vehicular Technology. His research interests include multiple access, wireless AI and RIS communications. He has published more than 200 papers in IEEE journals with citations more than11,000 in Google scholar.
\end{IEEEbiography}

\begin{IEEEbiography}[{\includegraphics[width=1in,height=1.25in,clip,keepaspectratio]{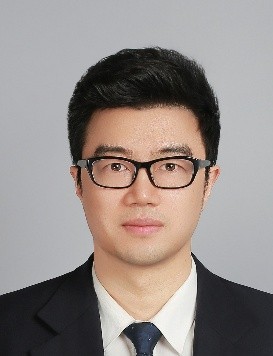}}]
{Jun Li}~(M’09-SM’16-F’25) received Ph.D. degree in Electronic Engineering from Shanghai Jiao Tong University, Shanghai, P. R. China in 2009. From January 2009 to June 2009, he worked in the Department of Research and Innovation, Alcatel Lucent Shanghai Bell as a Research Scientist. From June 2009 to April 2012, he was a Postdoctoral Fellow at the School of Electrical Engineering and Telecommunications, the University of New South Wales, Australia. From April 2012 to June 2015, he was a Research Fellow at the School of Electrical Engineering, the University of Sydney, Australia. From June 2015 to June 2024, he was a Professor at the School of Electronic and Optical Engineering, Nanjing University of Science and Technology, Nanjing, China. He is now a Professor at the School of Information Science and Engineering, Southeast University, Nanjing, China. He was a visiting professor at Princeton University from 2018 to 2019. His research interests include distributed intelligence, multiple agent reinforcement learning, and their applications in ultra-dense wireless networks, mobile edge computing, network privacy and security, and industrial Internet of Things. He has co-authored more than 300 papers in IEEE journals and conferences. He was serving as an editor of IEEE Transactions on Wireless Communication and TPC member for several flagship IEEE conferences.
\end{IEEEbiography}

\begin{IEEEbiography}[{\includegraphics[width=1in,height=1.25in,clip,keepaspectratio]{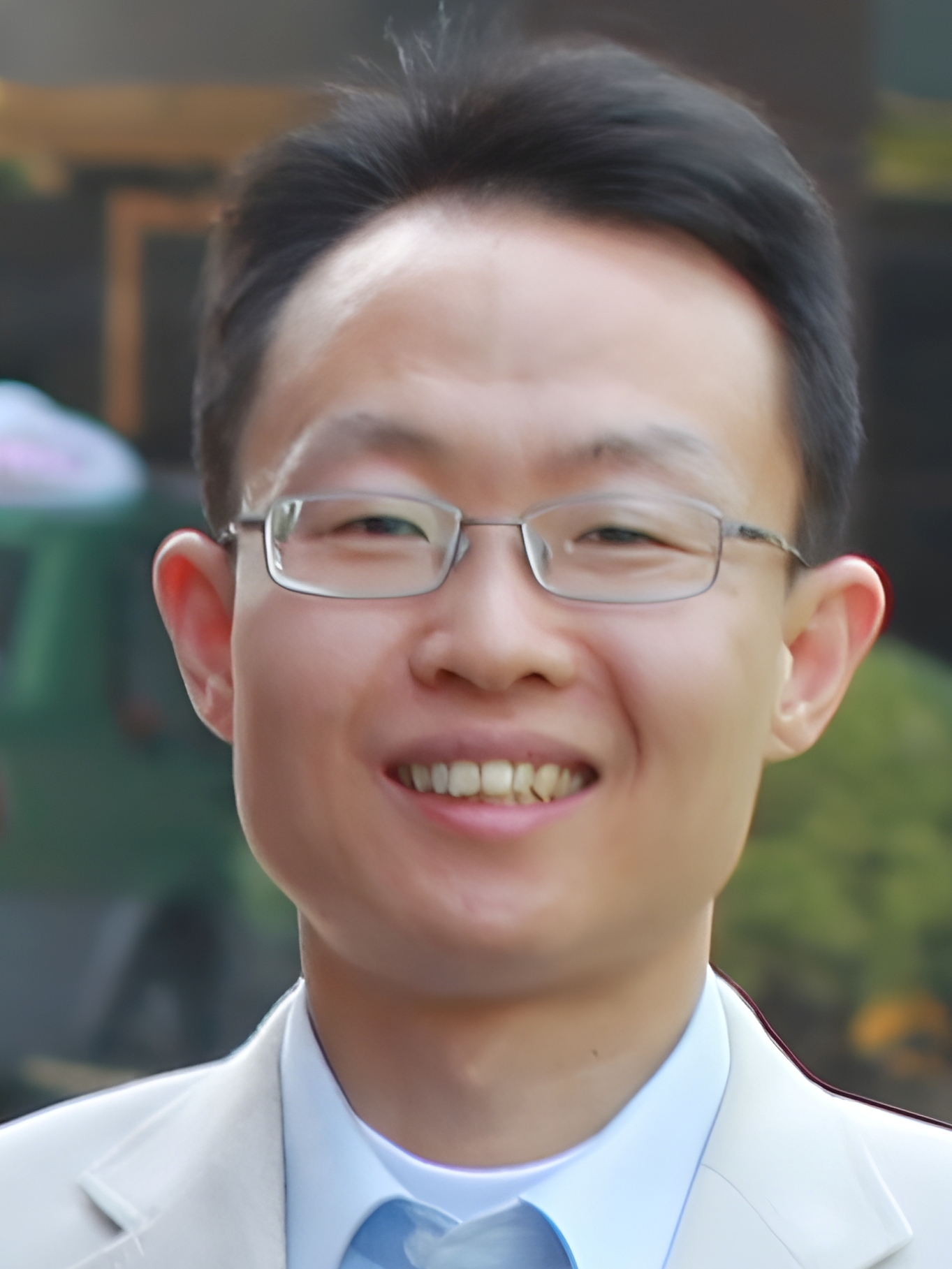}}]
{Ming Ding}~(IEEE M’12-SM’17) received the B.S. (with first-class Hons.) and M.S. degrees in electronics engineering from Shanghai Jiao Tong University (SJTU), China, and the Doctor of Philosophy (Ph.D.) degree in signal and information processing from SJTU, in 2004, 2007, and 2011, respectively. From April 2007 to September 2014, he worked at Sharp Laboratories of China as a Researcher/Senior Researcher/Principal Researcher. Currently, he is the Group Leader of the Privacy Technology Group at CSIRO’s Data61 in Sydney, NSW, Australia. Also, he is an Adjunct Professor at Swinburne University of Technology and University of Technology Sydney, Australia. His research interests include data privacy and security, machine learning and AI, and information technology. He has co-authored more than 300 papers in IEEE/ACM journals and conferences, all in recognized venues, and around 20 3GPP standardization contributions, as well as two books, i.e., ``Multi-point Cooperative Communication Systems: Theory and Applications'' (Springer, 2013) and “Fundamentals of Ultra-Dense Wireless Networks” (Cambridge University Press, 2022). Also, he holds 21 US patents and has co-invented another 100+ patents on 4G/5G technologies. Currently, he is an editor of IEEE Communications Surveys and Tutorials and IEEE Transactions on Network Science and Engineering. Besides, he has served as a guest editor/co-chair/co-tutor/TPC member for multiple IEEE top-tier journals/conferences and received several awards for his research work and professional services, including the prestigious IEEE Signal Processing Society Best Paper Award in 2022 and Highly Cited Researcher recognized by Clarivate Analytics in 2024.
\end{IEEEbiography}

\begin{IEEEbiography}[{\includegraphics[width=1in,height=1.25in,clip,keepaspectratio]{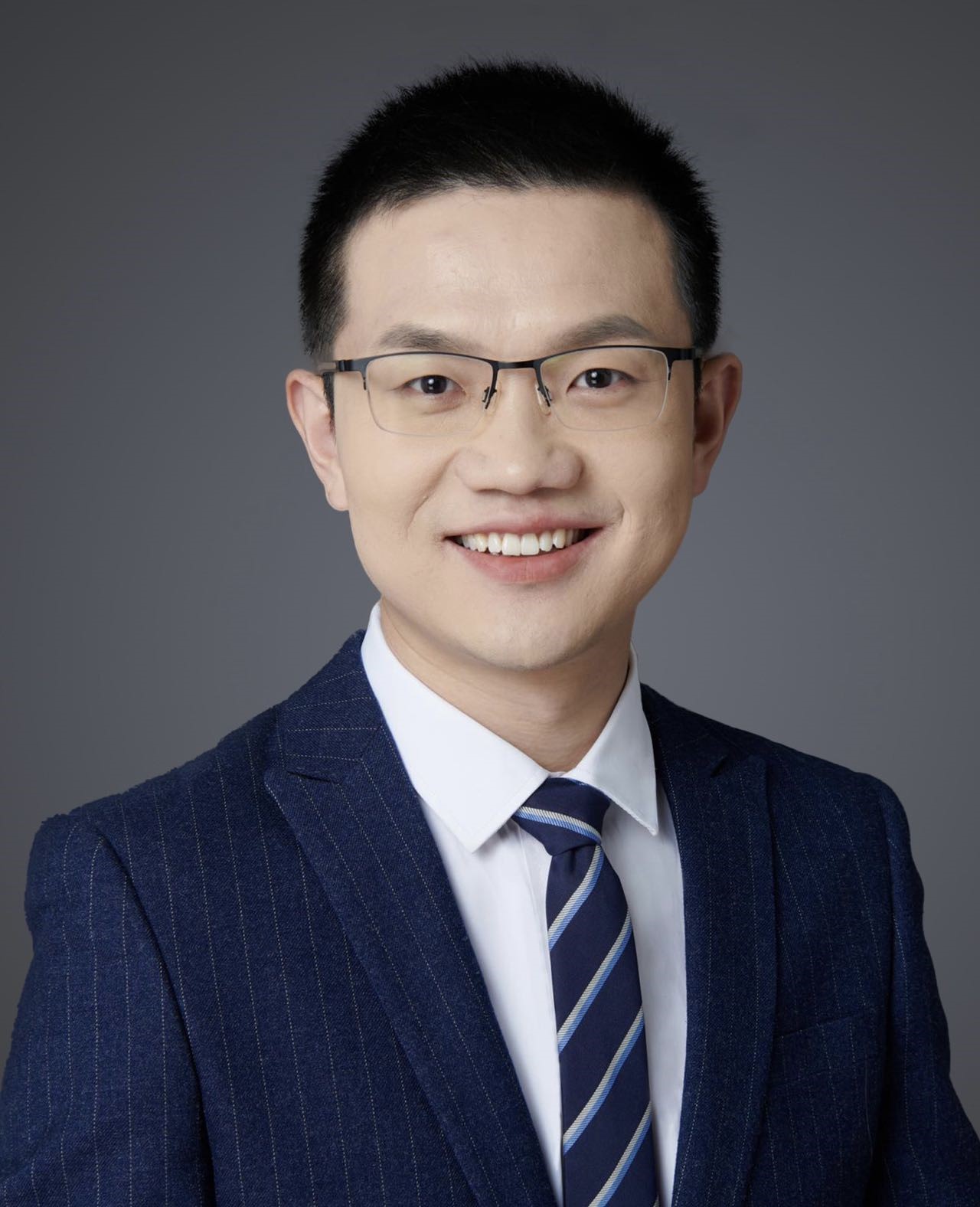}}]
{Qingqing Wu}~(S'13-M'16-SM'21) is an Associate Professor with Shanghai Jiao Tong University. His current research interest includes intelligent reflecting surface (IRS), unmanned aerial vehicle (UAV) communications, and MIMO transceiver design. He has coauthored more than 100 IEEE journal papers with 40+ ESI highly cited papers and 10+ ESI hot papers, which have received more than 31,000 Google citations. He has been listed as the Clarivate ESI Highly Cited Researcher since 2021, the Most Influential Scholar Award in AI-2000 by Aminer since 2021, World’s Top 2\% Scientist by Stanford University since 2020, and Xiaomi Young Scholar.

He was the recipient of the IEEE ComSoc Fred Ellersick Prize, Best Tutorial Paper Award in 2023, Asia-Pacific Best Young Researcher Award and Outstanding Paper Award in 2022, Young Author Best Paper Award in 2021 and 2024, the Outstanding Ph.D. Thesis Award of China Institute of Communications in 2017, the IEEE ICCC Best Paper Award in 2021, and IEEE WCSP Best Paper Award in 2015. He was the Exemplary Editor of IEEE Communications Letters in 2019 and the Exemplary Reviewer of several IEEE journals. He serves as an Associate/Senior/Area Editor for IEEE Transactions on Wireless Communications, IEEE Transactions on Communications, IEEE Communications Letters, IEEE Wireless Communications Letters. He is the Lead Guest Editor for IEEE Journal on Selected Areas in Communications. He is the workshop co-chair for IEEE ICC 2019-2023 and IEEE GLOBECOM 2020. He serves as the Workshops and Symposia Officer of Reconfigurable Intelligent Surfaces Emerging Technology Initiative and Research Blog Officer of Aerial Communications Emerging Technology Initiative. He has served as the Chair of the IEEE ComSoc Young Professional AP Committee and the Chair of IEEE VTS Drone Committee.
\end{IEEEbiography}
\vspace{-4mm}
\begin{IEEEbiography}[{\includegraphics[width=1in,height=1.25in,clip,keepaspectratio]{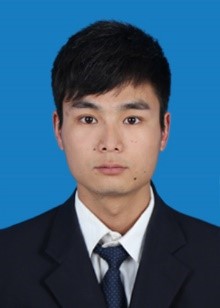}}]
{Kang Wei}~received his Ph.D. degree from Nanjing University of Science and Technology, Nanjing, China, in 2023. Before that, he received a B.S. degree in information engineering from Xidian University, Xian, China, in 2014. He is currently an associate professor at Southeast University. He has won the 2022 IEEE Signal Processing Society Best Paper Award and the 2022 Wiley China Open Science Author of the Year. He mainly focuses on privacy protection and optimization techniques for edge intelligence, including federated learning, differential privacy, and network resource allocation.
\end{IEEEbiography}
\vspace{-4mm}
\begin{IEEEbiography}[{\includegraphics[width=1in,height=1.25in,clip,keepaspectratio]{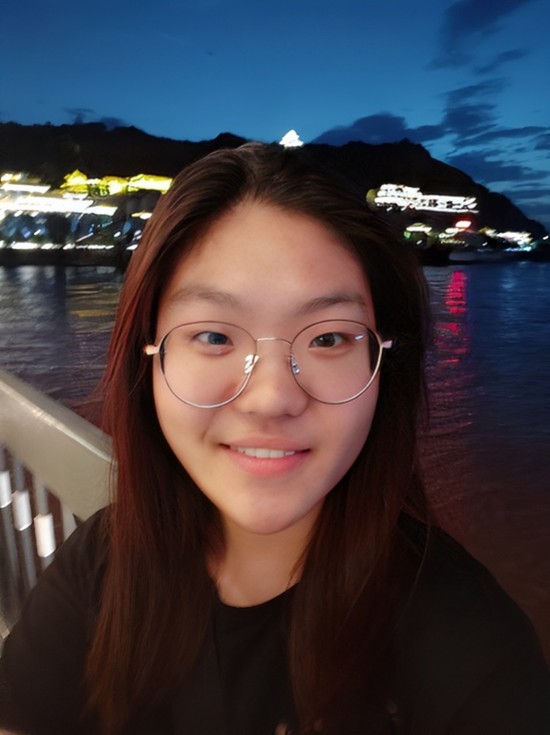}}]
{Xiumei Deng}~is currently a postdoctoral fellow at the Singapore University of Technology and Design. Before that, she received the Ph.D. degree in Information and Communications Engineering from Nanjing University of Science and Technology, China, in 2024, and the B.E. degree in Electronic Information Engineering from Nanjing University of Science and Technology, China, in 2018. Her research focuses on algorithm design and optimization for edge intelligence, with interests in edge generative AI, on-device large language models, federated learning, network resource optimization, and blockchain.
\end{IEEEbiography}
\vspace{-4mm}
\begin{IEEEbiography}[{\includegraphics[width=1in,height=1.25in,clip,keepaspectratio]{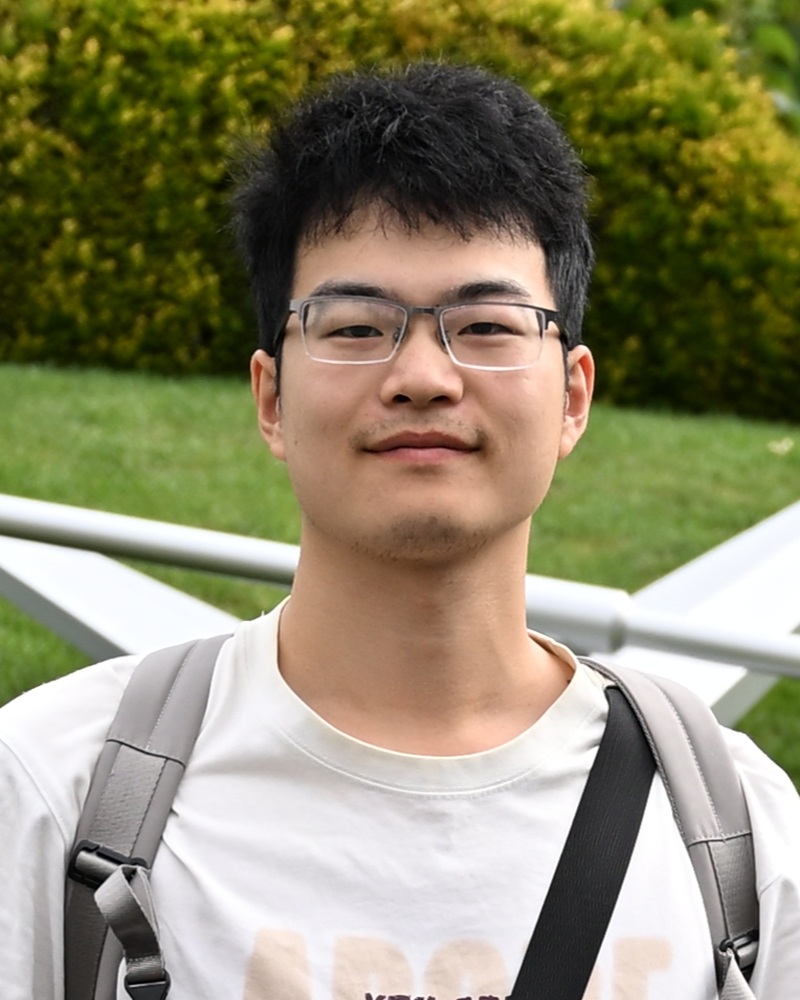}}]
{Yumeng Shao}~received the B.S. degree in communication engineering from the School of Electronic and Optical Engineering, Nanjing University of Science and Technology, Nanjing, China in 2019, and the Ph.D. degree in information and communication engineering from the School of Electronic and Optical Engineering, Nanjing University of Science and Technology, Nanjing, China. His research interests include distributed machine learning, blockchain, machine learning security and privacy, multimodal learning, and trusted AI.
\end{IEEEbiography}
\vspace{-4mm}
\begin{IEEEbiography}[{\includegraphics[width=1in,height=1.25in,clip,keepaspectratio]{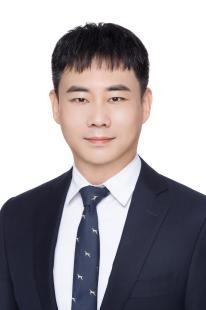}}]
{Qiong Wu}~(Senior Member, IEEE) is currently an associate professor with the School of Internet of Things Engineering, Jiangnan University, Wuxi, China. Dr. Wu is a Senior Member of IEEE and China Institute of Communications. He has published over 90 papers in high impact journals and conferences, and authorized over 30 patents. He was elected as one of the world's top 2\% scientists in 2024 and 2022 by Stanford University, and has received the young scientist award for ICCCS'24 and ICITE’24. He has been awarded the National Academy of Artificial Intelligence (NAAI) Certified AI Senior Engineer. He won the high-impact paper of Chinese Journal of Electronics award in 2024, and was the excellent reviewer for Computer Networks in 2024. He has severed as the (early career) editorial board member, (lead) guest editor, TPC co-chair, special session chair, workshop chair, TPC member and session chair for over 20 journals and conferences. His current research interest focuses on vehicular networks, autonomous driving communication technology, and machine learning.
\end{IEEEbiography}

\vfill\pagebreak

\end{document}